\documentclass[11pt]{article}

\title{\vspace{20mm} \textsl{On the BV formalism of open superstring field theory \\ in the large Hilbert space} \vspace{10mm} }
\author{\Large Hiroaki Matsunaga\footnote{matsunaga@fzu.cz \hspace{1mm} ${}^{\dagger }$nomura@hep1.c.u-tokyo.ac.jp \hspace{\fill } [UT-Komaba/18-1]} \vspace{1mm} } 
\date{${}^{\ast }$\textit{Institute of Physics of the Czech Academy of Sciences, \\ Na Slovance 2, Prague 8, Czech Republic} \\ \vspace{5mm}
{\Large Mitsuru Nomura${}^{\dagger }$} \\
\vspace{2mm} 
${}^{\dagger }$\textit{Institute of Physics, University of Tokyo, \\ Komaba, Meguro-ku, 153-8902, Japan} \\ \vspace{3mm} 
\vspace{0mm}}

\usepackage[dvips]{graphicx}
\usepackage{color, upgreek, mathrsfs, enumitem}
\usepackage{amsmath,amscd, amssymb, amsfonts, amsthm}

\usepackage[hypertex]{hyperref}
\usepackage{cite}

\makeatletter
  
  \@addtoreset{equation}{section}
\makeatother

\newcommand{\no}{\nonumber\\}

\newcommand{\ld}{ [ \hspace{-0.6mm} [ }
\newcommand{\rd}{ ] \hspace{-0.6mm} ] }
\newcommand{\Ld}{ \big[ \hspace{-1.1mm} \big[ }
\newcommand{\Rd}{ \big] \hspace{-1.09mm} \big] }
\newcommand{\LD}{ \Big[ \hspace{-1.3mm} \Big[ }
\newcommand{\RD}{ \Big] \hspace{-1.3mm} \Big] }
\newcommand{\la}{\big{\langle }}
\newcommand{\ra}{\big{\rangle }}
\newcommand{\La}{\Big{\langle }}
\newcommand{\Ra}{\Big{\rangle }}
\newcommand{\LA}{\bigg{\langle }}
\newcommand{\RA}{\bigg{\rangle }}
\newcommand{\lla}{\big{\langle } \hspace{-1.3mm} \big{\langle }} 
\newcommand{\rra}{\big{\rangle } \hspace{-1.3mm} \big{\rangle }}

\newcommand{\srra}{\rangle \hspace{-1mm} \rangle } 
\newcommand{\slla}{\langle \hspace{-1mm} \langle } 

\newcommand{\Eta}{\boldsymbol{\eta }} 
\newcommand{\bxi }{\boldsymbol{\xi }} 
 
\newcommand{\bLambda}{\boldsymbol{\Lambda }}

\newcommand{\bPhi }{\boldsymbol{\Phi }}

\newcommand{\bM}{\mathbf{M}}

\newcommand{\cA}{\mathcal{A}} 
\newcommand{\cB}{\mathcal{B}} 
 
\newcommand{\cD}{\mathcal{D}} 
\newcommand{\cE}{\mathcal{E}} 
\newcommand{\cF}{\mathcal{F}} 

\newcommand{\cH}{\mathcal{H}}

\newcommand{\cP}{\mathcal{P}} 
 
\newcommand{\cR}{\mathcal{R}} 
 
\newcommand{\cT}{\mathcal{T}}

\newcommand{\cZ}{\mathcal{Z}} 

\allowdisplaybreaks[3] 

\begin{document}
\maketitle\thispagestyle{empty}\addtocounter{page}{-1} 
\begin{abstract}
We construct several BV master actions for open superstring field theory in the large Hilbert space. 
First, we show that a naive use of the conventional BV approach breaks down at the third order of the antifield number expansion, although it enables us to define a simple ``string antibracket'' taking the Darboux form as spacetime antibrackets. 
This fact implies that in the large Hilbert space, ``string fields--antifields'' should be reassembled to obtain master actions in a simple manner. 
We determine the assembly of the string antifields on the basis of Berkovits' constrained BV approach, and give solutions to the master equation defined by Dirac antibrackets on the constrained string field--antifield space. 
It is expected that partial gauge-fixing enables us to relate superstring field theories based on the large and small Hilbert spaces 
directly: 
Reassembling string fields--antifields is rather natural from this point of view. 
Finally, inspired by these results, we revisit the conventional BV approach and construct a BV master action based on the minimal set of string fields--antifields. 
\end{abstract}
\clearpage {\small \tableofcontents } 

\section{Introduction}

Reducible gauge field theories require ghosts, ghosts-for-ghosts, and higher-ghosts as much as necessary, whose gauge algebras may necessitate equations of motion to be closed algebras. 
The Batalin-Vilkovisky (BV) formalism gives a convenient way to describe such gauge theories, in which the action is expressed in terms of fields and antifields \cite{bv papers, bv reviews}. 
A string field is an assembly of such spacetime fields and string field theory is an infinitely reducible gauge theory, in which the BV master action is expressed in terms of \textit{string fields} and \textit{string antifields} \cite{Witten:1985cc, bosonic bv, Zwiebach:1992ie, A infinity and bv}. 

\vspace{1mm} 

In bosonic string field theory, its classical BV master action is constructed by just relaxing the ghost number restriction of the classical action---there is a ready-made procedure. 
However, unfortunately, one cannot apply this procedure to superstring field theory in general; how to construct its BV master action has not been clarified yet, except for very limited cases. 
A successful formulation of open superstring field theory was first proposed by Berkovits \cite{Berkovits:1995ab}, which is characterized by a string field living in the \textit{large} Hilbert space \cite{Friedan:1985ge} and a Wess-Zumino-Witten-like (WZW-like) action having the large gauge invariance. 
Using a real parameter $t \in [0,1]$, this WZW-like action is often written in the following condensed form  
\begin{subequations}
\begin{align}
\label{S}
S [\Phi ] & = - \int_{0}^{1} dt \, \La \, A_{t} [ t \Phi ] \, , \, Q \, A_{\eta } [t \Phi ] \, \Ra _{\textsf{bpz}}
= - \frac{1}{2} \, \la \, \Phi \, , \, Q \, \eta \, \Phi  \, \ra _{\textsf{bpz}} + \cdots \, , 
\end{align}
where $Q$ is the BRST operator, $\eta $ denotes the zero-mode of the eta ghost, $\langle A , B \rangle _{\textsf{bpz}}$ is the BPZ inner product of $A,B$ in the large Hilbert space, and $A_{\eta } [\Phi ] = \eta \, \Phi + \cdots $ is a nonlinear functional\footnote{A functional $A_{t} [t \Phi ] = \partial_{t} ( t \Phi ) + \cdots $ is determined by given $A_{\eta } [\Phi ]$; for example, $A_{t} [t \Phi _{\textsf{B}}] = (\partial _{t} e^{t\Phi _{\textsf{B}}}) e^{-t \Phi _{\textsf{B}}} $. } of the dynamical string field $\Phi$ defined by a solution of the Maurer-Cartan equation 
\begin{align}
\label{pure}
0 & \equiv \eta \, A_{\eta } [ \Phi ] -  A_{\eta } [ \Phi ] \ast  A_{\eta } [ \Phi ] 
\, . 
\end{align}
The dots denote the nonlinear interacting terms. 
The symbol $\ast $ denotes Witten's associative star product \cite{Witten:1985cc}; using it, a solution of (\ref{pure}) is given by $A_{\eta } [ \Phi _{\textsf{B}} ] = (\eta \, e^{\Phi _{\textsf{B}}} ) e^{- \Phi _{\textsf{B}}}$. 
Since $Q$ and $\eta$ are nilpotent and graded commutative, it is invariant under the \textit{large} gauge transformations  
\begin{align}
\label{Phi}
\delta \Phi = Q \, \Lambda + \eta \, \Omega + \dots , 
\end{align}
\end{subequations} 
where $\Lambda$ and $\Omega$ are appropriate string fields of gauge parameters. 
This \textit{large} gauge symmetry complicates its gauge-fixing problem. 
A master action for the free theory can be constructed in a simple manner \cite{Torii:2011zz, Kroyter:2012ni, Torii:2012nj}; however, its ghost--antifield parts or BV transformations take somewhat different forms from the kinetic term of (\ref{S}) or the linear part of (\ref{Phi}), respectively. 
By contrast, a nonlinear BV master action has proven to be difficult to find even for perturbative one, and it has remained unsolved problem up until now. 

\vspace{1mm} 

There is a more practical option: 
One can ignore the large Hilbert space and consider superstring field theory within the small Hilbert space \cite{Erler:2013xta}; 
iff all fields and their gauge algebras are strictly restricted to be small, one can apply the ready-made BV procedure.\footnote{Higher string ghosts and their gauge algebra can be large keeping the dynamical string field small, in which the ready-made procedure is not applicable \cite{Matsunaga:2017phm}: It requires the BV formalism in the large Hilbert space. } 
Quantum aspects of superstring fields have been studied by utilizing such a gauge-fixable formulation. 
It is however expected that this gauge-fixable theory is obtained from superstring field theory in the large Hilbert space via partial gauge-fixing \cite{Iimori:2013kha, Iimori:2015aea, Erler:2015rra}.\footnote{Partial gauge fixing is an operation omitting some part $\Phi ^{\eta }$ of the dynamical field $\Phi = \Phi ^{\xi } + \Phi ^{\eta }$ \textit{by hand}; at the same time, corresponding gauge degrees of (\ref{Phi}) are appropriately omitted \textit{by hand}---like gauge fixing. } 
It seems to be intuitively clear at the classical level, but its validity has remained unclear because we have not succeeded to construct a BV master action in the large Hilbert space yet. 
A lack of understanding of the BV formalism of superstring field theory in the large Hilbert space is not just an issue for the Berkovits formulation but also a matter of all other formulations.\footnote{For including the Ramond sector, see \cite{Kunitomo:2015usa, Matsunaga:2015kra} for the large; see \cite{Erler:2016ybs, Konopka:2016grr} for the small. 
For closed superstring field theories, see \cite{Berkovits:2004xh, Matsunaga:2014wpa} for the large; see \cite{Erler:2014eba, Sen:2015uaa} for the small; see \cite{Goto:2015pqv} for intermediation. } 

\vspace{-1mm} 

\subsubsection*{BV master action in the large Hilbert space}  

Several differently-looking formulations of superstring field theory are currently established; these all can be embedded into the \textit{large} Hilbert space and their gauge structures are understood in a unified manner---at least, at the classical level. 
In this process, even for a very trivial embedding of a gauge-fixable theory based on the small Hilbert space, 
the original gauge symmetry is enlarged as (\ref{Phi}) and the Wess-Zumino-Witten-like gauge structure arises \cite{Matsunaga:2017phm, Erler:2015uoa, Matsunaga:2016zsu, Erler:2017onq}. 
Since the gauge-fixing of WZW-like string field theory has been an unsolved problem, this result implies that a gauge-fixable theory turns into a gauge-unfixable theory after the embedding---so, the BV master action should exist even in the large Hilbert space. 
To see it, we consider the simplest situation, the \textit{large} $A_{\infty }/L_{\infty }$ theory: An action in the large Hilbert space which is defined by the trivial embedding of a gauge-fixable action in the small Hilbert space \cite{Erler:2015rra, Erler:2015uoa}. 

\vspace{1mm} 

In this paper, we construct BV master actions for open superstring field theory in the large Hilbert space on the basis of several different approaches. 
Through these constructions, we would like to clarify the following questions; ``How can we apply the BV formalism to the \textit{large} theory?'', ``Why does our ready-made BV procedure not work in the large Hilbert space?'' and ``What should we take into account to treat \textit{large} gauge symmetries?'', which are our motivations. 
In the most of this paper, we focus on the Neveu-Schwarz (NS) sector of open superstring field theory, the \textit{large} $A_{\infty }$ theory. 
One can apply the completely same prescriptions to the Ramond sector or to the \textit{large} $L_{\infty }$ action for closed superstring field theory. 

\vspace{-1mm} 

\subsubsection*{As an example of the WZW-like action} 

The large $A_{\infty }$ theory is the simplest but a nontrivial example of the WZW-like open string field theory \cite{Erler:2017onq}: 
It is completely described by a pair of mutually commutative $A_{\infty }$ algebras $(\Eta \,; \bM )$; 
alternatively, it is also described by the equivalent $A_{\infty }$ pair $(\Eta - \boldsymbol{\ast } \,; \mathbf{Q} )$. 
The large $A_{\infty }$ action just equals to the WZW-like action (\ref{S}) based on another solution of (\ref{pure}) given by 
\begin{align}
\label{another sol}
A_{\eta } [ t \Phi ] \equiv \pi _{1} \widehat{\textbf{G}} \frac{1}{1 - t \, \eta \, \Phi } \, , 
\end{align}
where $\widehat{\mathbf{G}}$ is an $A_{\infty }$ morphism satisfying $\widehat{\mathbf{G}} \, \Eta = (\Eta - \boldsymbol{\ast }) \widehat{\mathbf{G}}$ and $\widehat{\mathbf{G}} \, \bM = \mathbf{Q} \, \widehat{\mathbf{G}}$ given in \cite{Erler:2015rra, Erler:2015uoa}. 
Hence, our construction of BV master actions for the large $A_{\infty }$ theory provides evidence of BV master actions for the WZW-like formulation; it will be a first step to clarify the BV formalism of WZW-like superstring field theory in the large Hilbert space. 

\vspace{-1mm} 

\subsubsection*{Organization of the article}

To apply the BV formalism, we have to analyse the gauge reducibility of the theory and find the minimal set of the fields--antifields, which we first explain in section 2. 
We also explain that the conventional BV approach provides an elegant string field representation of the BV antibracket, which is very useful for perturbative constructions. 
In section 3, we show that although one can construct a lower order BV master action up to the second order of the antifield number, there exists no proper solution at higher order within the (naive) conventional BV approach. 
Section 4 is devoted to explain how we can avoid the no-go result of section 3 by using the constrained BV approach \cite{Batalin:1992mk}. 
In this approach, the construction of a master action is equivalent to specify the form of an (unconstrained) action $S_{\textsf{bv} }$ and constraints $\widehat{\Gamma }$. 
In addition to it, we have to specify how to assemble \textit{extra string fields} $\varphi _{\textrm{ex}}$ in string field theory. 
So, we have to find out an appropriate pair $( S_{\textsf{bv}} , \widehat{\Gamma } , \varphi _{\textrm{ex} } )$ giving a proper solution of the constrained BV master equation, which we explain in section 5. 
First, we show that Berkovits' prescription \cite{Berkovits:2012np} works well and gives a correct constrained BV master action for \textit{partially gauge-fixed} superstring field theory in the large Hilbert space. 
In order to remove this partially-gauge-fixing assumption, one has to impose further constraints, reassemble the extra string fields, or replace the starting (unconstrained) BV action. 
Then, we construct appropriate constrained BV actions in the large Hilbert space (without the partially-gauge-fixing assumption) on the basis of several different prescriptions. 
In particular, a constrained BV master action obtained in section 5.4 resembles canonical transformations switching $Q$- and $\eta$-gauge symmetries \cite{Matsunaga:2017phm}, and these properties may be helpful to see what happens in the large theory. 
In section 6, we revisit the conventional BV approach. 
On the basis of remediations inspired by the results of the constrained BV approach, we construct a BV master action within the conventional BV approach. 
Notations, basic identities, and some elementary facts are in appendix A. 

\section{Minimal set: String fields--antifields}

In this paper, we clarify how to apply the BV formalism to superstring field theory in the large Hilbert space by using the large $A_{\infty }$ theory---the simplest example of the WZW-like formulation. 
As we will see, the classical action of the large $A_{\infty }$ theory 
\begin{subequations} 
\begin{align}
\label{original action a}
S [\Phi ] = - \frac{1}{2} \la \, \Phi \, , \, Q \, \eta \, \Phi \, \ra _{\textsf{bpz}}
- \frac{1}{3} \la \, \Phi \, , \, M_{2} \big{(} \eta \, \Phi , \eta \, \Phi \big{)} \ra _{\textsf{bpz}}
- \frac{1}{4} \la \, \Phi , \, M_{3} \big{(} \eta \, \Phi , \eta \, \Phi , \eta \, \Phi \big{)} \ra _{\textsf{bpz}}
+ \cdots 
\end{align} 
is given (or defined) by the trivial embedding of the \textit{small} $A_{\infty }$ theory. 
In this section, we analyse its gauge reducibility and give the minimal set of string fields--antifields. 

Let $\Phi $ be a Neveu-Schwarz (NS) open superstring field living in the large Hilbert space, which carries world-sheet ghost number $0$ and picture number $0$. 
The \textit{large} string field $\Phi $ reduces to a \textit{small} string field $\Psi \equiv \eta \, \Phi $ by acting $\eta$ on it; 
the \textit{small} string field $\Psi$ satisfies $\eta \, \Psi = 0$ and carries world-sheet ghost number $1$ and picture number $-1$, which lives in the small Hilbert space. 
We write $\bM = \mathbf{Q} + \bM _{2} + \cdots $ for the NS open superstring products given by \cite{Erler:2013xta}: 
The $g$-th product $M_{g}$ carries world-sheet ghost number $2-g$ and picture number $g-1$. 
As a functional of the \textit{small} dynamical string field $\Psi$, the \textit{small} $A_{\infty }$ action $S' [\Psi ]$ is given by 
\begin{align*}
& S' [\Psi ] = - \frac{1}{2} \lla \, \Psi \, , \, Q \, \Psi \, \rra _{\textsf{bpz}}
- \frac{1}{3}  \lla \, \Psi \, , \, M_{2} ( \Psi , \Psi )  \rra _{\textsf{bpz}}
- \frac{1}{4} \lla \, \Psi \, , \, M_{3} \big{(} \Psi , \Psi , \Psi \big{)} \rra _{\textsf{bpz}}
+ \cdots \, . 
\end{align*}
We write $\slla \eta A, \eta B \srra _{\textsf{bpz}}$ for the BPZ inner product of $\eta A$ and $\eta B$ in the small Hilbert space, which equals to the BPZ inner product $\langle A , \eta B \rangle _{\textsf{bpz}} = -(-)^{A} \langle \eta A , B \rangle _{\textsf{bpz}}$ in the large Hilbert space. 
This small $A_{\infty }$ theory is easily gauge-fixable iff all string fields of gauge parameters and their gauge algebras are also restricted to the small Hilbert space: 
One can construct its BV master action $S'_{\textsf{bv}}$ by just relaxing ghost number constraint as $S'_{\textsf{bv}} \equiv S' [\psi ]$ where $\psi $ carries all spacetime and world-sheet ghost numbers. 
By contrast, one cannot construct the BV master action $S_{\textsf{bv}}$ for the large $A_{\infty }$ action (\ref{original action a}) in a similar manner because of its WZW-like large gauge symmetries, although $S[\Phi ]$ is obtained from the trivial embedding of gauge-fixable $S'[\Psi ]$, which we explain. 

\subsection{Gauge reducibility and string ghosts}

For simplicity, we take coalgebraic and suspended notation; see appendix A. 
With a real parameter $t \in [0,1]$, the large $A_{\infty }$ action (\ref{original action a}) has the following compact expression 
\begin{align}
\label{original action}
S [\Phi ] 
& = \int_{0}^{1} dt \, \La \, \Phi \, , \, \bM \frac{1}{1- t \, \eta \, \Phi } \, \Ra \, , 
\end{align}
\end{subequations} 
where $\bM = \mathbf{Q} + \bM _{2} + \bM _{3} + \cdots $ denotes the $A_{\infty }$ superstring products and $\langle A , B \rangle $ is the graded symplectic form---it is the suspended BPZ inner product (\ref{susp}), but we call it as ``the BPZ inner product'' simply. 
This action is invariant under the following large gauge transformations 
\begin{align}
\label{gauge invariance}
\delta \Phi = \pi _{1} \Ld \bM , \, \bLambda _{-1,0} \Rd \frac{1}{1-\eta \, \Phi } + \Eta \, \Lambda _{-1,1} \, , 
\end{align}
where $\ld \mathbf{C} , \mathbf{D} \rd$ denotes the graded commutator of coderivations $\mathbf{C}$ and $\mathbf{D}$; see appendix A. 
The gauge symmetry (\ref{gauge invariance}) has the following gauge reducibility 
\begin{align*}
\delta _{g+1} \big( \delta _{g} \Lambda _{-g,p} \big) = 0 \, , \hspace{5mm} 
\delta _{g} \Lambda _{-g,p}  = \pi _{1} \Ld \bM , \, \bLambda _{-g-1,p} \Rd \frac{1}{1-\eta \, \Phi } + \Eta \, \Lambda _{-g-1,p+1} \, , 
\end{align*}
where $\Lambda _{-g,p}$ denotes a $g$-th string gauge-parameter field and defining $\Lambda _{0,0} \equiv \Phi $ may be helpful. 
While the $g$-label runs from $0$ to infinity, the $p$-label runs from $0$ to $g$ as shown by \cite{Torii:2011zz, Kroyter:2012ni, Torii:2012nj}. 
Hence, the large $A_{\infty }$ theory is infinitely reducible just as the Berkovits theory \cite{Berkovits:1995ab}. 

\vspace{1mm} 

As well as the string field $\Phi $, these string gauge-parameter fields $\Lambda _{-g,p}$ can be expanded in terms of spacetime gauge-parameter fields $\lambda _{g,p}^{\, r}$ and world-sheet bases $| \cZ _{-g,p}^{\, r} \rangle $; see appendix A for these bases. 
The BV formalism implies that when string gauge-parameter fields are given by $\Lambda _{-g,p} = \sum_{r} \lambda _{-g,p}^{\, r} | \cZ _{-g,p}^{\, r} \rangle$, corresponding \textit{string ghost fields} are obtained by replacing each spacetime parameter field $\lambda _{-g,p}^{\, r}$ with corresponding spacetime ghosts $\phi _{-g,p}^{\, r}$ as follows 
\begin{align}
\label{ghost string field}
\Phi _{-g,p} = \sum_{r} \phi _{g,p}^{\, r} \, \big{|} \cZ _{-g,p}^{\, r} \ra \, . 
\end{align}
The $r$-label distinguishes different bases carrying the same world-sheet ghost and picture numbers. 
We sometimes write $\Phi _{0,0} \equiv \Phi $ for simplicity. 
Therefore, the following set of the dynamical string field $\Phi $ and string ghosts $\Phi _{-g,p}$ arises in the theory 
\begin{align}
\label{string ghost set}
\Big{\{ } \, \Phi _{-g,p} \, \Big{|} \, \Phi _{0,0} \equiv \Phi \, ;  \, 0 \leq g \, , \,\, 0 \leq p \leq g \, \Big{\} }  \, . 
\end{align}
More precisely, all of the spacetime fields which are coefficients of these string fields (\ref{string ghost set}) are necessitated to fix the gauge symmetries (\ref{gauge invariance}). 
In other words, a set of spacetime ghost fields $A'_{\textrm{min}} \equiv \{ \, \phi _{g,p}^{r} \, | \, 0<g , \, 0 \leq p \leq g \, ; r \in \mathbb{N} \, \}$ are required. 
We write $A_{0}$ for the set of spacetime dynamical fields, $A_{0} \equiv \{ \phi _{0,0}^{r} \}_{r\in \mathbb{N}}$. 
The pair $A_{\textrm{min}} \equiv A_{0} \oplus A'_{\textrm{min}}$ of dynamical fields $A_{0}$ and these ghosts $A'_{\textrm{min}}$ requires their spacetime antifields $A_{\textrm{min}}^{\ast } = \{ (\phi _{g,p}^{r})^{\ast } | \, 0 \leq g , \, 0 \leq p \leq g \, ; r \in \mathbb{N} \, \}$ in the BV formalism. 
Hence, the minimal set of spacetime fields--antifields is given by 
\begin{align}
\label{minimal set}
\cA _{\textrm{min}} \equiv A_{\textrm{min}} \oplus A_{\textrm{min}}^{\ast } 
= \Big{\{ } \, \phi _{g,p}^{r} \, , \, (\phi _{g,p}^{r} )^{\ast } \, \Big{|} \, 0 \leq g \, , \,\, 0 \leq p \leq g \,\, ; \, r \in \mathbb{N} \, \Big{\} }  \, . 
\end{align}
On this minimal set, one can define a non-degenerate antibracket  
\begin{align}
\label{minimal antibracket}
\big{(} \, F \, , \, G \, \big{)}_{\textrm{min}} 
\equiv \sum_{g \geq 0} \sum_{p, r} \bigg[ \, 
\frac{\overset{\leftarrow }{\partial } F}{\partial \phi _{g,p}^{\, r} } \, 
\frac{\overset{\rightarrow }{\partial } G}{\partial (\phi _{g,p}^{\, r})^{\ast } } \, 
- \, 
\frac{\overset{\leftarrow }{\partial } F}{\partial (\phi _{g,p}^{\, r})^{\ast } } \, 
\frac{\overset{\rightarrow }{\partial } G}{\partial \phi _{g,p}^{\, r} } \, 
\bigg] \, , 
\end{align}
where $\frac{\overset{\rightarrow }{\partial } F}{\partial \phi }$ is the left-derivative, $\frac{\overset{\leftarrow }{\partial } F}{\partial \phi }$ is the right-derivative, and $\frac{\overset{\rightarrow }{\partial } F}{\partial \phi } = (-)^{\phi (F+1)} \frac{\overset{\leftarrow }{\partial } F}{\partial \phi }$ holds. 
One can quickly find $( F ,G )_{\textrm{min}} = -(-)^{(F+1)(G+1)} ( G , F )_{\textrm{min}}$ in this expression.

\subsection{String antifields and string antibracket}

In the conventional BV approach, for a given string (ghost) field $\Phi _{-g,p}$ of (\ref{ghost string field}), its string antifield $(\Phi _{-g,p})^{\ast }$ is introduced by assigning $|\cZ _{-g,p}^{\, r \, \ast }\rangle $ to each spacetime antifield $( \phi _{g,p}^{\, r} )^{\ast }$, 
\begin{align}
\label{naive string antifield}
(\Phi _{-g,p})^{\ast } = \sum_{r} (\phi _{g,p}^{\, r})^{\ast } \, \big{|} \cZ _{-g,p}^{\, r \, \ast } \ra \, , 
\end{align}
where $|\cZ _{-g,p}^{\, r \, \ast }\rangle $ is the dual basis for $|\cZ _{-g,p}^{\, r}\rangle $ such that $\langle \cZ _{-g,p}^{\, r \, \ast } , \cZ _{-h,q}^{\, s} \rangle = - \delta ^{r,s} \, \delta _{g,h} \, \delta _{p,q}$\,. 
Since dual bases are uniquely determined for given bases, this type of string antifield seems to be most natural. 
See appendix A for details of these bases. 
The set of string fields (\ref{string ghost set}) and their string antifields defined by (\ref{naive string antifield}) gives the minimal set of \textit{string fields--antifields} 
\begin{align}
\label{string fields--antifields} 
\cA _{\Phi } |_{\textrm{min}} \equiv \Big{\{ } \, \Phi _{-g,p} \, , (\Phi _{-g,p} )^{\ast } \, \Big{|} \, 0 \leq g  \, , \, 0 \leq p \leq g \, \Big{\} } \, .
\end{align}
As shown by \cite{Kroyter:2012ni}, this type of string antifield (\ref{naive string antifield}) provides an elegant string field representation of the BV antibracket (\ref{minimal antibracket}) in the Darboux form, 
\begin{align}
\label{string field rep}
\big{(} \, F \, ,\, G \, \big{)} _{\textrm{min}} 
= \sum_{g,p} \bigg[ \, 
F \frac{\overset{\leftarrow }{\partial } }{\partial \Phi _{-g,p}} \cdot \frac{\overset{\rightarrow }{\partial } }{\partial (\Phi _{-g,p})^{\ast }} G  
- F \frac{\overset{\leftarrow }{\partial } }{\partial (\Phi _{-g,p})^{\ast }} \cdot \frac{\overset{\rightarrow }{\partial } }{\partial \Phi _{-g,p}} G \, 
\bigg] \, , 
\end{align}
where $A \cdot B$ denotes the BPZ inner product in the large Hilbert space $A \cdot B \equiv \langle A , B \rangle$\,. 
Note that $F$ and $G$ are functionals of (\ref{string fields--antifields}), which can be identified with functionals of (\ref{minimal set}). 
One can define \textit{string-field derivatives} of a functional $F=F[\Phi _{\alpha }]$ of string fields $\Phi _{\alpha }$ by utilizing its total derivative $\delta F $. 
When a given string field $\Phi _{\alpha }$ consists of spacetime fields $\{ \phi _{\alpha }^{r} \} _{r}$ as $\Phi _{\alpha } = \sum_{r} \phi _{\alpha }^{r} | \cZ_{\alpha }^{r} \rangle $, we require that the total derivative of $F = F[\Phi _{\alpha }] = F [\{ \phi _{\alpha }^{r} \} _{r}]$ has the following expression 
\begin{align*}
\LA \, \delta \Phi _{\alpha } \, , \, \frac{\overset{\rightarrow }{\partial } F}{\partial \Phi _{\alpha }} \, \RA 
\equiv 
\sum_{r} \delta \phi _{\alpha }^{r} \, \frac{\overset{\rightarrow }{\partial } }{\partial \phi _{\alpha }^{r}} F \, , 
\hspace{5mm}  
\LA \, \frac{\overset{\leftarrow }{\partial } F}{\partial \Phi _{\alpha }} \, , \, \delta \Phi _{\alpha } \, \RA 
\equiv 
F \sum_{r} \frac{\overset{\leftarrow }{\partial } }{\partial \phi _{\alpha }^{r}} \, \delta \phi _{\alpha }^{r} \, , 
\end{align*} 
which provides the string-field derivatives. 
Note that the relation of the left- and right-derivatives $\frac{\overset{\rightarrow }{\partial } F}{\partial \phi } = (-)^{\phi (F+1)} \frac{\overset{\leftarrow }{\partial } F}{\partial \phi }$ determines that of string-field derivatives. 
We assume that the variations of string fields (\ref{ghost string field}) and string antifields (\ref{naive string antifield}) are given by 
\begin{align*}
\delta \Phi _{-g,p} \equiv \sum_{r} \delta \phi_{g,p}^{r} \, \big{|} \cZ _{-g,p}^{\, r} \ra \, , 
\hspace{5mm} 
\delta (\Phi _{-g,p})^{\ast } = \sum_{r} \delta (\phi _{g,p}^{r})^{\ast } \, \big{|} \cZ _{-g,p}^{\, r \, \ast } \ra \, . 
\end{align*} 
Then, by using the relations (\ref{1>g}) and (\ref{g>1}), the BV antibracket (\ref{minimal antibracket}) reduces to (\ref{string field rep}). 

\vspace{1mm} 

Let us consider the free action $K[\Phi ]$ and its gauge variation---the kinetic term of (\ref{S}) and the linear part of (\ref{Phi}). 
Its master action $K_{\textsf{bv}}$ gives the kinetic term of the master action $S_{\textsf{bv}}$. 
As shown in \cite{Kroyter:2012ni}, a master action for the free theory is given by 
\begin{align} 
\label{free}
K_{\textsf{bv}}[\Phi , \Phi ^{\ast }] = \frac{1}{2} \la \, \Phi \, , \, Q \, \eta \, \Phi \, \ra 
+ \sum_{g \geq 0} \sum_{p=0}^{g} \la ( \Phi _{-g,p} )^{\ast } , \, Q \, \Phi _{-1-g,p} + \eta \, \Phi _{-1-g,p+1} \, \ra \, , 
\end{align} 
which is indeed a functional of string fields (\ref{ghost string field}) and string antifields (\ref{naive string antifield}).

\section{Conventional BV approach}

In the conventional BV approach, we require the following three properties to obtain a master action $S_{\textsf{bv}} = S_{\textsf{bv}} [\varphi , \varphi ^{\ast }]$ as a functional of \textit{string fields} $\varphi $ and \textit{string antifields} $\varphi ^{\ast }$: 

\begin{enumerate}[label=\textbf{\roman*)}, leftmargin=!]
\item Regarding the states: The master action $S_{\textsf{bv}}$ consists of the dynamical string field, the string ghost fields introduced in (\ref{ghost string field}), and the string antifields given by (\ref{naive string antifield}). 
\item Regarding the operators and products: The master action $S_{\textsf{bv}}$ consists of the operators and products which appear in the original action (\ref{original action}) and its gauge symmetry algebra (\ref{gauge invariance}), namely, $\bM$, $\Eta$, and the large BPZ inner product only. 
\item The master action $S_{\textsf{bv}}$ does not include explicit insertions of $\bxi $ or $\bM ^{-1}$: These operations enable us to remove the above requirement (i) or (ii) effectively. 
\end{enumerate}

However, although a perturbative master action $S_{\textsf{bv}} = S^{(0)} + S^{(1)} + S^{(2)} + \dots $ is obtained up to the second order, this (naive) conventional BV approach breaks down at the third order $S^{(3)}$ of the antifield number expansion. 
There is no solution satisfying the above three requirements, which we explain in this section. 
A reader interested in constructing BV master actions can skip this section; 
this section is independent of the other sections. 

\vspace{1mm} 

As expected, if one use $\bxi $ or $\bM ^{-1}$ insertions explicitly, a master action can be constructed.\footnote{Then, some of higher gauge tensors have to include $\bxi $ or $\bM ^{-1}$ explicitly, although it does not appear in (\ref{original action}) or (\ref{gauge invariance}) explicitly. 
The results of section 5 and 6 imply that there may be difference between the gauge tensors based on string fields--antifields and those based on spacetime fields--antifields. } 
It implies that string ghost fields or string antifields are reassembled, or new products which never appear in the action nor its gauge invariance are used, to obtain the master action. 
In section 5, keeping the forms of string ghost fields and the requirement for operators and products, we construct the master action by just reassembling (physical) string antifields. 

\subsection{Naive construction breaks down} 

We perturbatively solve the master equation using the antifield number expansion. 
We write $\textrm{afn}[\phi ]$ for the antifield number of the spacetime field or antifield $\phi$. 
It is assigned to the spacetime antifields only: $\textrm{afn}[ \phi ]=0$ if $\phi $ is not an antifield. 
In particular, $\textrm{afn}[c]=0$ for $c \in \mathbb{C}$ and a world-sheet basis has no antifield number. 
The antifield number is additive with respect to the multiplication $\textrm{afn}[\phi \psi ]=\textrm{afn}[\phi ] + \textrm{afn}[\psi ]$, and thus $\textrm{afn}[\phi ] + \textrm{afn}[ \frac{\partial }{\partial \phi }]=0$\,. 
We find 
\begin{align*} 
\textrm{afn} [ \phi _{g,p} ] = 
- \textrm{afn} \bigg[ \frac{\overset{\rightarrow }{\partial } }{\partial \phi _{g,p}} \bigg] = 0 
\, , \hspace{3mm} 
\textrm{afn} \big[ (\phi _{g,p} )^{\ast } \big] = - \textrm{afn} \bigg[ \frac{\overset{\rightarrow }{\partial } }{\partial (\phi _{g,p})^{\ast }} \bigg] = 
g + 1 \, , 
\end{align*}
where $\phi _{g,p}$ denotes a $g$-th ghost. 
A master action $S_{\textsf{bv}}$ is a functional of all fields--antifields appearing in the minimal set and one can expand it with respect to the antifield number 
\begin{align*}
S_{\textsf{bv}} = S + \sum_{a=1}^{\infty } S^{(a)} \, ,  
\end{align*}
where $S^{(a)}$ denotes the antifield number $a$ part of the master action $S_{\textsf{bv}}$, namely, $\textrm{afn}[S^{(a)}] = a$\,. 
The original action is the antifield number zero part $S^{(0)} \equiv S$, which is the initial condition of the BV formalism. 
Because of $\textrm{afn} \big[ \frac{\partial S^{(a)} }{\partial (\phi _{g,p} )^{\ast } } \big] = a-g-1$ and $\textrm{afn} \big[ \frac{\partial S^{(a)} }{\partial \phi _{g,p} } \big] = a$\,, the antifield number $a$ part of the master equation is given by 
\begin{subequations} 
\begin{align} 
\label{a-part of BV eq} 
\frac{1}{2} \big{(} \, S_{\textsf{bv}} \, , \, S_{\textsf{bv}} \, \big{)} \big{|}^{(a)}_{\textrm{min}} 
\equiv \sum_{b=0}^{a} \sum_{g=0}^{b} \bigg[ \sum_{p=0}^{g} 
S^{(a-[b-g])} \frac{\overset{\leftarrow }{\partial }}{\partial \phi _{g,p} } 
\frac{\overset{\rightarrow }{\partial }}{\partial (\phi _{g,p})^{\ast } } S^{(1+b)} \bigg] 
= 0 \, . 
\end{align}
Note that $(S_{\textsf{bv}} , S_{\textsf{bv}})|_{\textrm{min}}^{(a)}$ consists of $S^{(0)},  \cdots , S^{(a+1)}$ because of $\frac{\partial S^{(a)}}{\partial \phi _{g,p}}= \frac{\partial S^{(a)}}{\partial (\phi _{g,p})^{\ast }} = 0$ for $a \leq g$\,. 
By solving these, one can construct a solution $S_{\textsf{bv}}$ of the master equation 
\begin{align*} 
\big{(} \, S_{\textsf{bv}} \, , \, S_{\textsf{bv}} \, \big{)}_{\textrm{min}} 
= \sum_{a=0}^{\infty } 
\big{(} \, S_{\textsf{bv}} \, , \, S_{\textsf{bv}} \, \big{)} \big{|}^{(a)}_{\textrm{min}} = 0 \, . 
\end{align*}
In the conventional BV approach, (\ref{a-part of BV eq}) has the following string field representation 
\begin{align} 
\frac{1}{2} \big{(} \, S_{\textsf{bv}} \, , \, S_{\textsf{bv}} \, \big{)} \big{|}^{(a)}_{\textrm{min}} 
\equiv \sum_{b=0}^{a} \sum_{g=0}^{b} \bigg[ \sum_{p=0}^{g} 
S^{(a-[b-g])} \LA \frac{\overset{\leftarrow }{\partial }}{\partial \Phi _{-g,p} } \, , \,  
\frac{\overset{\rightarrow }{\partial }}{\partial (\Phi _{-g,p})^{\ast } } \RA S^{(1+b)} \bigg]  
= 0 \, . 
\end{align}
\end{subequations}
Note that the antifield number expansion of $S_{\textsf{bv}}$ defines the following odd vector field $\overset{\rightarrow }{\Delta }$ lowering the antifield number by one, $\textrm{afn}[\overset{\rightarrow }{\Delta }]=-1$, 
\begin{align*}
\overset{\rightarrow }{\Delta } \equiv \sum_{g=0}^{\infty } \sum_{p=0}^{g} S^{(g)} 
\frac{\overset{\leftarrow }{\partial }}{\partial \phi _{g,p} } \, 
\frac{\overset{\rightarrow }{\partial }}{\partial (\phi _{g,p})^{\ast } }
= \sum_{g=0}^{\infty } \sum_{p=0}^{g} 
\frac{\overset{\leftarrow }{\partial } S^{(g)} }{\partial \Phi _{-g,p} } \cdot 
\frac{\overset{\rightarrow }{\partial }}{\partial (\Phi _{-g,p})^{\ast } } \, , 
\end{align*}
where the dot denotes the BPZ inner product in the large Hilbert space. 
The odd vector field $\overset{\rightarrow }{\Delta }$ acting on $S^{(a+1)}$ is uniquely determined by given lower parts $S^{(0)}, \cdots , S^{(a)}$. 
The first equation $( S_{\textsf{bv}} , S_{\textsf{bv}} )|^{(0)}_{\textrm{min}} = 0$ reduces to $\overset{\rightarrow }{\Delta } S^{(1)} = 0$ because of $\frac{\partial S^{(a)}}{\partial (\phi _{g,p})^{\ast }}=0$ for $a \leq g$\,; 
a solution is given by 
\begin{align}
\label{S^{(1)}}
S^{(1)} = \La \, (\Phi )^{\ast } \, , \, \textrm{M} (\Phi _{-1,0} ) + \eta \, \Phi _{-1,1} \Ra \, , 
\end{align}
where we wrote $\textrm{M} ( \Phi _{0} ) \equiv \pi _{1} \, \Ld \bM , \bPhi _{0} \Rd \frac{1}{1-\eta \, \Phi }$ for brevity. 
The antifield number 2 part $S^{(2)}$ is determined by the second equation $(S^{(1)} , S^{(1)})_{\textrm{min}} + \overset{\rightarrow }{\Delta } S^{(2)} = 0$\,. 
To be proper, $S^{(2)}$ has to include $\Phi _{-2,p}$ and $(\Phi _{-1,p})^{\ast }$ in addition to $\Phi _{-1,p}$ and $(\Phi )^{\ast }$. 
We find a solution 
\begin{align}
\label{S^{(2)}}
S^{(2)} & = \La \, (\Phi _{-1,0} )^{\ast } \, , \, 
\textrm{M} ( \Phi _{-2,0} ) + \frac{1}{2} \textrm{M} ( \Phi _{-1,0} , \eta \, \Phi _{-1,0} ) 
+ \eta \, \Phi _{-2,1} \Ra \, 
\no & \hspace{5mm} 
+ \La (\Phi _{-1,1} )^{\ast } \, , \, 
\textrm{M} ( \Phi _{-2,1} )  
- \frac{1}{2} \textrm{M} \big{(} \Phi _{-1,0} , \textrm{M} ( \Phi _{-1,0} ) \big{)} 
+ \eta \, \Phi _{-2,2} \Ra \, 
\no & \hspace{5mm} 
+ \La (\Phi )^{\ast } \, , \, 
\frac{1}{2} \textrm{M} \big{(} \Phi _{-2,0} , ( \Phi )^{\ast } \big{)} 
+ \frac{1}{4} \textrm{M} \big{(} \Phi _{-1,0} , \eta \, \Phi _{-1,0} , ( \Phi )^{\ast } \big{)} \Ra \, , 
\end{align}
where we defined $\textrm{M} ( \Phi _{1} , \Phi _{2} ) \equiv \pi _{1} \, \Ld \ld \bM , \bPhi _{1} \rd , \bPhi _{2} \Rd \frac{1}{1-\eta \, \Phi } = (-)^{\Phi _{1} \Phi _{2} } \textrm{M} (\Phi _{2} , \Phi _{1} )$ for brevity. 
The second line includes a double $\bM$-term, and $\bM$-terms and $\Eta$-terms appear in symmetric manner. 
We introduce the following graded symmetric function of $n$-inputs 
\begin{align*}
\textrm{M} ( \Phi _{1} , \dots , \Phi _{n} ) & \equiv \pi _{1} \, \LD ... \Ld \ld \bM , \bPhi _{1} \rd , \bPhi _{2} \Rd , \dots \Rd , \bPhi _{n} \RD \frac{1}{1-\eta \, \Phi } \, , 
\end{align*}
which satisfies $\textrm{M}( ... , A , B , ... ) = (-)^{AB} \textrm{M} ( ... , B , A , ... )$\,. 

As (\ref{S^{(1)}}) and (\ref{S^{(2)}}), we would like to construct the next correction $S^{(3)}$ satisfying the antifield number $2$ part of the master equation (\ref{a-part of BV eq}). 
However, there is no solution based on the (naive) conventional BV approach: 
One cannot construct higher $S^{(a \geq 3)}$ as a functional of $\Phi _{-g,p}$ and $(\Phi _{-g,p})^{\ast }$ unless using projectors acting on string fields--antifields.\footnote{For a given string field $\varphi $, we split it as $\varphi = \varphi _{1} + \cdots + \varphi _{n}$; for each split-part $\varphi _{a}$, we introduce its (split) string antifield $(\varphi _{a} )^{\ast }$, which may satisfy $\varphi ^{\ast } = (\varphi _{1})^{\ast } + \cdots + (\varphi _{n})^{\ast }$. 
As we will see in section 5 or 6, the master action $S_{\textsf{bv}}$ can be constructed as a functional of these split-parts $\varphi _{1} , ... , \varphi _{n}$ of the string (anti-)field, $S_{\textsf{bv}} = S_{\textsf{bv}} [ \varphi _{a} , (\varphi _{a})^{\ast } ]$. 
It is not a functional of the sum $\varphi = \varphi _{1} + \cdots + \varphi _{n}$ or $\varphi ^{\ast } = (\varphi _{1} )^{\ast } + \cdots + (\varphi _{n} )^{\ast }$. 
So we need $\cP _{a}$ s.t. $\cP _{a} \varphi = \varphi _{a}$.}  
The above lower order solutions $S^{(1)}$ and $S^{(2)}$ uniquely determine the following quantity, 
\begin{align*} 
F ^{(2)} & \equiv  
\LA \frac{ \overset{\leftarrow }{\partial } S^{(2)} }{\partial \Phi } \, , \,  
\frac{\overset{\rightarrow }{\partial } S^{(1)} }{\partial (\Phi )^{\ast } } \RA  
+ 
\LA \frac{\overset{\leftarrow }{\partial } S^{(1)} }{\partial \Phi } \, , \, 
\frac{\overset{\rightarrow }{\partial } S^{(2)} }{\partial (\Phi )^{\ast } } \RA 
+
\sum_{p=0}^{1} 
\LA \frac{\overset{\leftarrow }{\partial } S^{(2)} }{\partial \Phi _{-1,p} } \, , \,  
\frac{\overset{\rightarrow }{\partial } S^{(2)} }{\partial (\Phi _{-1,p})^{\ast } } \RA \, . 
\end{align*} 
The antifield number 2 part of the master equation (\ref{a-part of BV eq}) is equivalent to 
\begin{align}
\label{broken eq}
F^{(2)} + \overset{\rightarrow }{\Delta } \, S^{(3)} = 0 \, . 
\end{align}
The odd vector field $\overset{\rightarrow }{\Delta }$ acting on $S^{(3)}$ is uniquely determined by $S^{(0)}$, $S^{(1)}$, and $S^{(2)}$. 
Note that a proper $S^{(3)}$ must include $\Phi _{-3,p}$ and $(\Phi _{-2,p})^{\ast }$. 
The equation (\ref{broken eq}) should hold for each pair of string field variables, and one can find solutions for generic pairs; however, the equation (\ref{broken eq}) has no solution for the pair of string field variable $(\Phi _{-1,0} , \Phi _{-1,0} , \Phi _{-1,0} , (\Phi _{-1,1})^{\ast })$ unfortunately. 
We find that $\overset{\rightarrow }{\Delta } S^{(3)}$ has the following form, 
\begin{align*}
\overset{\rightarrow }{\Delta } S^{(3)} = \LA (\textrm{e.o.m.}) \, , \, 
\frac{\overset{\rightarrow }{\partial } S^{(3)} }{\partial (\Phi )^{\ast } } \RA 
+ \LA \eta \, (\Phi _{-1,1} )^{\ast } , \, 
\frac{\overset{\rightarrow }{\partial } S^{(3)} }{\partial (\Phi _{-2,2})^{\ast } } \RA  
+ \LA \textrm{M} \big{(} ( \Phi _{-1,1} )^{\ast } \big{)} , \, 
\frac{\overset{\rightarrow }{\partial } S^{(3)} }{\partial (\Phi _{-2,1})^{\ast } } \RA 
+ \cdots , 
\end{align*}
where the last dots denote the terms which consist of the other pairs of variables. 
The corresponding terms of $F^{(2)}$ must be able to be rewritten into the same form to satisfy (\ref{broken eq}). 
Unfortunately, we find 
\begin{align*}
F^{(2)} = \La \, (\textrm{e.o.m.}) , \,  F \, \Ra 
+ \La \, \eta \, (\Phi _{-1,1} )^{\ast } , \, F_{\eta } \, \Ra 
+ \La \, \textrm{M} \big{(} \Phi _{-1,1} )^{\ast } \big{)} , \, F_{\textrm{M}} \, \Ra 
+ \La ( \Phi _{-1,1})^{\ast } , \, E \, \Ra 
+ \dots \, , 
\end{align*}
where the dots denote the terms which consist of the other pairs of variables and the explicit forms of $F$, $F_{\textrm{M}}$, $F_{\eta }$, and $E$ are given by 
\begin{align*}
F & \equiv - \frac{1}{4} \Big[ \textrm{M} \Big{(} \Phi _{-1,0} , \eta \, \Phi _{-1,0} , \textrm{M} \big{(} \Phi _{-1,0} , (\Phi _{-1,1})^{\ast } \big{)} \Big{)} 
+ \textrm{M} \Big{(} \Phi _{-1,0} , \textrm{M} \big{(} \Phi _{-1,0} , \eta \, \Phi _{-1,0} , (\Phi _{-1,1})^{\ast }  \big{)} \Big{)} \Big] \, , 
\\ 
F_{\eta } & \equiv - \frac{1}{4} \Big[ \textrm{M} \big{(} \Phi _{-1,0} , \textrm{M} (\Phi _{-1,0}) , \textrm{M} (\Phi _{-1,0})  \big{)}  
+ \textrm{M} \big{(} \Phi _{-1,0} , \eta \, \Phi _{-1,0} , \textrm{M} (\Phi _{-1,0} ) \big{)} \Big] \, , 
\\ 
F_{\textrm{M}} & \equiv \frac{1}{4} \textrm{M} \big{(} \Phi _{-1,0} , \eta \, \Phi _{-1,0} , \textrm{M} ( \Phi _{-1,0} ) \big{)} \, , 
\hspace{15mm}  
E \equiv \frac{1}{4} \textrm{M} \big{(} \Phi _{-1,0} , \textrm{M} ( \Phi _{-1,0} , \eta \, \Phi _{-1,0} ) \big{)} \, . 
\end{align*}
The nonzero fourth term, which is extra and breaks (\ref{broken eq}), cannot be absorbed by the first three terms. 
Hence, although one can construct a lower order solution $S_{\textsf{bv}} = S + S^{(1)} + S^{(2)} + O(3)$, there is no solution for higher $S^{(a>2)}$ based on the (naive) conventional BV approach.

\subsection{On the gauge tensor formulae} 

The BV master equation is a generating function of the identities satisfied by the gauge tensors---what does the breakdown of (\ref{broken eq}) mean? 
Let us consider the gauge tensors arising from (\ref{gauge invariance}). 
We write $\cR^{\alpha }_{\beta }$, $\cT ^{\alpha }_{\beta \gamma }$, or $\cE ^{\alpha \beta }_{\gamma \delta }$ for gauge tensors in the sense of 
\begin{align*}
\delta \Phi = \cR ^{\Phi }_{\alpha } (\Lambda _{\alpha } ) \, , 
\hspace{3mm} 
\ld \delta _{1} , \delta _{2} \rd \Phi 
= \cR ^{\Phi }_{\gamma } \, \cT ^{\gamma }_{\alpha _{1} \alpha _{2} } ( \Lambda _{\alpha _{1}}, \Lambda _{\alpha _{2}} ) 
- \frac{\partial S}{\partial \Phi '} \, \cE ^{\Phi ' \Phi }_{\alpha _{2} \alpha _{1}} (\Lambda _{\alpha _{1}} , \Lambda _{\alpha _{2}} ) \, , 
\end{align*}
where $\cR ^{\Phi } \equiv \cR ^{(0,0)}$ and the Greek indices denote appropriate world-sheet ghost and picture numbers: $\cR ^{\Phi }_{\alpha } (\Lambda _{\alpha } ) = \cR ^{\Phi }_{(-1,0)}( \Lambda _{-1,0} ) + \cR ^{\Phi }_{(-1,1)} (\Lambda _{-1,0} )$. 
These $\cR $, $\cT$, and $\cE$ define the following gauge tensors, which include terms corresponding to $(\Phi _{-1,0} , \Phi _{-1,0} , \Phi _{-1,0} , (\Phi _{-1,1})^{\ast })$, 
\begin{align*}
\cA _{\alpha \beta \gamma }^{\delta } 
& \equiv \frac{1}{3} \sum _{\textrm{cuclic}} \bigg[ \frac{\partial \cT ^{\delta }_{\alpha \beta } }{\partial \Phi '} \, \cR ^{\Phi '}_{\gamma } 
- \cT ^{\delta }_{\alpha \iota } \, \cT ^{\iota }_{\beta \gamma }  \bigg] \, . 
\\
\cB ^{\delta \iota }_{\alpha \beta \gamma } 
& \equiv \frac{1}{3} \sum_{\textrm{cyclic}} \bigg[ \frac{\partial \cE ^{\delta \iota } }{\partial \Phi '} \, \cR ^{\Phi '}_{\gamma } 
- \cE ^{\delta \iota }_{\alpha \beta } \, \cT^{\delta }_{\beta \gamma } 
- \frac{\partial \cR ^{\delta }_{\alpha }}{\partial \Phi '} \, \cE ^{\Phi' \iota }_{\beta \gamma } 
+ \frac{\partial \cR ^{\iota }_{\alpha }}{\partial \Phi ' } \, \cE^{\Phi ' \delta }_{\beta \gamma }  \bigg] \, . 
\end{align*}
We find the following relation of the on-shell Jacobi identity of the gauge transformations
\begin{align*}
\cR ^{i }_{\delta } \, \cA ^{\delta } _{\alpha \beta \gamma } = \frac{\partial S}{\partial \Phi '} \, \cB ^{\Phi ' \delta }_{\alpha \beta \gamma } \, 
\hspace{3mm} \Longleftrightarrow \hspace{3mm} 
\sum_{\textrm{cyclic}} \Ld \ld \delta _{1} , \delta _{2} \rd , \delta _{3} \Rd \Phi 
= 3 \frac{\partial S}{\partial \Phi '} \, \cB ^{\Phi ' \Phi }_{\alpha _{3} \alpha _{2} \alpha _{1} } \big{(} \Lambda _{\alpha _{1} } , \Lambda _{\alpha _{2} } , \Lambda _{\alpha _{3} } \big{)} \, . 
\end{align*}
Then, the master equation (\ref{broken eq}) is equivalent to the existence of the higher gauge tensors $\cF _{\alpha \beta \gamma }^{\delta }$ and $\cD _{\alpha \beta \gamma }^{\delta \iota }$ satisfying 
\begin{align}
\label{gauge tensor eq}
\cA ^{\delta }_{\alpha \beta \gamma } 
= - \cR ^{\delta }_{\iota } \, \cF ^{\iota }_{\alpha \beta \gamma } + \frac{\partial S}{\partial \Phi '} \, \cD ^{\Phi ' \delta }_{\alpha \beta \gamma } \, . 
\end{align} 
The right-hand side has the form as $\overset{\rightarrow }{\Delta }S^{(3)}$ of (\ref{broken eq}); the left-hand side provides the same kind of extra terms as $F^{(2)}$ of (\ref{broken eq}). 
Note however that the relation (\ref{gauge tensor eq}) should hold automatically when the set of independent gauge generators is complete. 
Since these gauge tensors $\cR$, $\cT$, $\cE$, ... are naively defined as functionals of \textit{string fields} $\varphi $, this result implies that one should have to consider them as functionals of \textit{spacetime fields} or rather fine parts $\varphi _{a}$ of total string fields $\varphi = \varphi _{1} + \cdots + \varphi _{n}$. 
The master action $S_{\textsf{bv}}$ will consist of these fine gauge tensors.

\section{Nonminimal set: Constrained string fields--antifields}

The previous no-go result implies that we should not consider $S_{\textsf{bv}}$ as a functional of string fields--antifields naively. 
We should consider rather fine parts $\varphi _{a}$ of string fields--antifields $\varphi = \sum_{a} \varphi _{a}$ such as spacetime fields; or equivalently, we have to reassemble the string fields--antifields in order to obtain $S_{\textsf{bv}}$ as a functional of string fields--antifields themselves. 

\subsection{How to assemble string antifields} 

There is no criteria or rule for how to assemble string antifields unlike string ghost fields in the BV formalism: 
It just suggests that how or what kind of spacetime ghost fields must be introduced from the gauge invariance, and one can introduce their spacetime antifields such that the antibracket takes the Darboux form. 
It just tells us whether a given master action, a functional of \textit{spacetime} fields--antifields, is proper or not. 
In general, the string antifield $(\Phi _{g,p})^{\ast }$ for the string field $\Phi _{g,p}$ can take the following form,  
\begin{align}
\label{p-relax}
(\Phi _{g,p} )^{\ast } = \sum_{r} (\phi _{g,p}^{r} )^{\ast } \, \big{|} \, g ,  p \, ; r \, \ra \, , \hspace{5mm} 
\big{|} \, g,p \, ; r \, \ra = \sum_{h,q} (a_{g,p}^{r})_{h,q} \, \big{|} \cZ _{2+h,-1+q}^{\, r} \ra \, ,
\end{align}
where $(a_{g,p}^{r})_{h,q}$ is some constant. 
As we saw in section 2, the relation $\langle \Phi _{g,p} , (\Phi _{g,p})^{\ast } \rangle =1$ gives the simplest assembly. 
For generic assembly of string antifields, its ``string field representation'' of the antibracket cannot take the Darboux form: 
When $(a_{g,p}^{r})_{h,p} \not= 0$ for $h \not= 0$ or $q \not= 0$, we find $( F ,G )_{\textrm{min}} = \sum F \overset{\leftarrow}{\partial }_{a} E^{ab} \overset{\rightarrow }{\partial }_{b} G$ where $\partial _{a}$ denotes a string (anti-)field derivative and $E^{ab}$ is not an orthogonal antisymmetric matrix. 
In this paper, we consider the case $(a_{g,p}^{r})_{h,p}=0$ for $h \not= 0$ but may be $(a_{g,p}^{r})_{0,p} \not= 0$, which depends on the construction. 
Then, all components of $E^{ab}$ have the same Grassmann parity, but its string antibracket may not be Darboux for the $p$-label.\footnote{The spacetime antibracket can always take the Darboux form even if its string field representation cannot. } 
 
In the large Hilbert space, one can split a given state into its $\eta $- and $\xi $-exact components as 
\begin{subequations} 
\begin{align}
\label{decomposition}
\Phi _{-g,p} = \sum_{r} \phi _{g,p}^{r \, \eta } \, \eta \, \xi  \, \big{|} \cZ _{-g,p}^{\, r} \ra + \sum_{r} \phi _{g,p}^{r \, \xi } \, \xi \, \eta \,  \big{|} \cZ _{-g,p}^{\, r} \ra \, . 
\end{align}
The new label of $\phi ^{ \eta}$ or $\phi ^{\xi }$ denotes that it is a coefficient spacetime field multiplied by an $\eta$- or $\xi $-exact world-sheet basis, respectively. 
Inspired by the formal relation $\langle \Phi _{g,p} , \, (\Phi _{g,p})^{\ast } \rangle \not= 0$, we require that the string antifield $(\Phi _{-g,p})^{\ast }$ for (\ref{decomposition}) 
takes the following form 
\begin{align}
(\Phi _{-g,p} )^{\ast } =  
\sum_{r} ( \phi _{g,p}^{r \, \xi } )^{\ast } \, \eta \, \xi  \, \big{|} \, g,p \, ; r \, \ra
+ \sum_{r} ( \phi _{g,p}^{r \, \eta } )^{\ast } \, \xi \, \eta \,  \big{|} \, g,p \, ; r \, \ra \, .
\end{align}
In other words, the $\eta $-exact components $(\Phi ^{\ast })^{\eta }$ of the string antifield $\Phi ^{\ast }= (\Phi ^{\ast })^{\eta } + (\Phi ^{\ast })^{\xi }$ correspond to the $\xi $-components $\Phi ^{\xi }$ of the string field $\Phi = \Phi ^{\eta } + \Phi ^{\xi }$ because of $\langle \Phi ^{\eta } , (\Phi ^{\ast })^{\eta } \rangle = \langle \Phi ^{\xi } , (\Phi ^{\ast })^{\xi } \rangle = 0$\,. 
In terms of spacetime fields, we assume 
\begin{align}
\big{(} (\phi _{g,p} )^{\ast } \big{)}^{\eta } = \big{(} \phi _{g,p}^{\,\, \xi } \big{)}^{\ast } \, , 
\hspace{5mm} 
\big{(} (\phi _{g,p} )^{\ast } \big{)}^{\xi } = \big{(} \phi _{g,p}^{\,\, \eta } \big{)}^{\ast } \, . 
\end{align}
\end{subequations} 
As will see, this requirement simplifies our analysis and computations. 

\vspace{1mm}

How should we assemble the string antifields?---to find it, we take the constrained BV approach, in which a solution of the constrained master equation determines how the physical string antifields should be assembled. 
We first consider extra string fields as \cite{Berkovits:2012np} and introduce their string antifields as the (naive) conventional BV approach. 
Then, we impose appropriate second constraints and consider the Dirac antibracket, which determines the physical string antifields on the constrained state space---appropriately assembled string antifields. 
In other words, we translated finding the assembly of string antifields into finding out appropriate constraints such that its Dirac antibracket is well-defined and the constrained master action becomes the generator of appropriate BV transformations. 
As will be discussed in 5.3, for example, one can introduce string antifield $(\Phi _{-g,p})^{\ast }|_{\Gamma }$ for (\ref{decomposition}) via appropriate constraints $\Gamma $ as 
\begin{align}
\label{example}
(\Phi _{-g,p} )^{\ast } \big{|}_{\Gamma } 
= \sum_{r} \eta \, \bigg[ \big{(}\phi _{g,p}^{r \, \xi } \big{)}^{\ast } \big{|} \cZ _{1+g ,-p}^{\, r} \ra 
+ \big{(} \phi _{g,p}^{r \, \eta } \big{)}^{\ast } \big{|} \cZ _{1+g, -1-p}^{\, r} \ra \bigg] \, , 
\end{align}
which gives a master action on the constrained string field--antifield space. 
Then, physical string antifields $\widetilde{\Phi } = \widetilde{\Phi } [ \Phi , \Phi ^{\ast } ; \Gamma ]$ are given by a funtcional of these string fields--antifields which satisfy the canonical relation $( \Phi _{a} , \widetilde{\Phi }_{b} )_{\Gamma } = \delta _{a,b}$ in the Dirac antibracket defined by (\ref{Dirac antibracket}).

\subsection{Extra fields and constraints} 

The Berkovits' constrained BV approach \cite{Berkovits:2012np} is a specific case of the BV formalism based on a redundant non-minimal set \cite{Batalin:1992mk}. 
As many as the antifields included in the minimal set, we introduce \textit{extra} spacetime ghost fields which carry negative spacetime ghost number, 
\begin{align}
\label{spacetime extra fields}
A_{\textrm{ex}} = \Big{\{ } \, \phi _{-1-g,-p}^{\, r} \, \Big{|} \,\, 0 \leq g \, , \,\, 0 \leq p \leq g \,\, ; \, r \in\mathbb{N} \, \Big{\} } \, . 
\end{align} 
For these extra spacetime ghosts, we introduce their spacetime antifields $A_{\textrm{ex}}^{\ast} = \{ (\phi _{-g,-p} )^{\ast } |\, 0 < g , 0 \leq p \leq g \}$. 
These give a set of extra fields--antifields $\cA _{\textrm{ex}} = A_{\textrm{ex}} \oplus A_{\textrm{ex}}^{\ast }$\,. 
We consider the \textit{non-minimal set} of fields--antifields 
\begin{align} 
\label{nonminimal set}
\cA = \cA _{\textrm{min}} \oplus \cA _{\textrm{ex}} 
= \Big{\{ } \phi _{g,p}^{r} \, , \, (\phi _{g,p}^{r} )^{\ast } \, ; \phi _{-1-g,-p}^{\, r} \, , \, (\phi _{-1-g,-p}^{\, r} )^{\ast} \, \Big{|} \, 0 \leq p \leq g \,\, ; r \in \mathbb{N} \, \Big{\} }  \, , 
\end{align}
and define an antibracket acting on this $\cA$ by 
\begin{align}
\label{antibracket}
\big{(} \, F \, , \, G \, \big{)} 
\equiv \sum_{g \in \mathbb{Z}} \sum_{p,r} \bigg[ \, 
\frac{\partial _{r} F}{\partial \phi _{g,p}^{r} } \, \frac{\partial _{l} G}{\partial \phi _{g,p}^{r \, \ast } }  
- \frac{\partial _{r} F}{\partial \phi _{g,p}^{r \, \ast } } \, \frac{\partial _{r} G}{\partial \phi _{g,p}^{r} } \, \bigg] \, . 
\end{align}
We introduce a set of constraint equations $\Gamma $ which has the same degrees of freedom as a set of extra fields--antifields $\cA _{\textrm{ex}}$\,. 
These $\Gamma $ split into the first and second class constraints. 
For any functions $F,G$ which are invariant under the first class $\Gamma$, one can define a non-degenerate Dirac antibracket on $\cA / \Gamma $ using the second class $\Gamma $ by 
\begin{align}
\label{Dirac antibracket}
\big{(} \, F \, , \, G \, \big{)}_{\Gamma } \equiv \big{(} \, F \, , \, G \, \big{)} - \sum_{a,b} \big{(} \, F \, , \, \Gamma _{a} \, \big{)} \, \big{[} ( \Gamma , \Gamma )^{-1} \big{]}_{ab} \, \big{(} \, \Gamma _{b} \, , \, G \, \big{)} \, . 
\end{align} 
It enables us to consider a master equation on the constraint space $\cA / \Gamma $. 
See \cite{Batalin:1992mk} for details. 

\vspace{1mm} 

In this paper, we just consider extra spacetime ghosts which have the same labels as original spacetime antifields appearing in the minimal set as \cite{Berkovits:2012np}. 
Hence, our non-minimal set of fields--antifields (\ref{nonminimal set}) is twice the size of the minimal set (\ref{minimal set}). 
We construct constrained master actions $S_{\textsf{bv}}$ based on this redundant set (\ref{nonminimal set}) and constrained brackets (\ref{Dirac antibracket}), instead of (\ref{minimal set}) and (\ref{minimal antibracket}).
In string field theory, by using these extra spacetime fields, we can introduce \textit{extra string fields}  
\begin{align*}
\Phi _{g,-p} = \sum_{r} \phi _{-g,p}^{r} \, \big{|} \, g  , p \, ; r \, \ra \,  
\end{align*} 
and consider a set of \textit{extra} string fields--antifields. 
In principle, there is no restriction on its assembly as long as it gives a solution of the constrained BV master equation.

\section{Constrained BV approach} 

Let us consider the ghost string fields $\Phi _{-g,p}$ of (\ref{ghost string field}), which are naturally determined by the gauge reducibility. 
For each spacetime field $\phi _{g,p}^{r}$ of $\Phi _{-g,p} = \sum _{r} \phi _{g,p}^{r} \, | \cZ _{-g,p}^{r} \rangle $\,, its spacetime antifield $(\phi _{g,p}^{r})^{\ast }$ is introduced. 
Now we add extra ghosts and their antifields, and our set of fields--antifields (\ref{nonminimal set}) is non-minimal and twice the size of (\ref{minimal set}). 
The pair of spacetime fields and their antifields $\{ \phi _{g,p}^{r} , (\phi _{g,p}^{r} )^{\ast } \} _{g,p,r}$ defines a non-degenerate antibracket (\ref{antibracket}) on functions of fields--antifields. 

\vspace{2mm} 

We consider a set $\{ \Phi _{1+g,-p} | 0 \leq g , \,\, 0 \leq p \leq g \}$ of \textit{extra string fields}, string fields consisting of extra spacetime ghosts, and assume that as string ghost fields (\ref{ghost string field}), these are assembled as 
\begin{align}
\label{extra ghost string field}
\Phi _{1+g,-p} = \sum_{r} \phi _{-1-g,-p}^{r} \, \big{|} \cZ _{1+g,-p}^{r} \ra \, . 
\end{align} 
This type of extra string field has the \textit{same} Grassmann parity or total grading as original string fields. 
As will see, $\{ \eta \, \Phi _{1+g,-p} \} _{g,p}$ will correspond to a half of conventional string antifields. 

\vspace{2mm} 

Let $\cA _{\varphi } \equiv \{ \Phi _{-g,p} , \Phi _{1+g,-p} \} _{0 \leq p \leq g}$ be the set of \textit{all string fields}\,: the dynamical string field, string ghost fields, and extra string fields. 
We write $\varphi $ for the sum of all string fields, 
\begin{align}
\label{all string fields}
\varphi \equiv \varphi _{-} + \varphi _{\textsf{ex}} \, , \hspace{5mm} 
\varphi _{-} \equiv \Phi + \sum _{g > 0} \sum_{p=0}^{g} \Phi _{-g,p} \, , \hspace{3mm} 
\varphi _{\textsf{ex}} \equiv \sum _{g \geq 0} \sum_{p=0}^{g} \Phi _{1+g,-p}  \, , 
\end{align}
where $\varphi _{-}$ denotes the sum of the original string fields and $\varphi _{\textsf{ex}}$ denotes the sum of the extra string fields. 
As proposed by Berkovits \cite{Berkovits:2012np}, we take the following constrained BV action 
\begin{align}
\label{Berkovits BV}
S_{\textsf{bv}} [\varphi ] = \int _{0}^{1} dt \, \La \, \varphi \, , \, \bM \frac{1}{1 - t \, \eta \, \varphi } \, \Ra \, , 
\end{align}
which has the same form as the original action (\ref{original action}). 
Clearly, this $S_{\textsf{bv}}[\varphi ]$ is not proper on (\ref{nonminimal set}) and satisfies $( S_{\textsf{bv}} , S_{\textsf{bv}} ) = 0$ because it consists of fields only. 
We introduce antifields into $S_{\textsf{bv}}$ such that it gives a proper master action by imposing appropriate constraints $\widehat{\Gamma }$. 

Note that as well as the original action $S[\Phi ]$, the above action $S_{\textsf{bv}}[\varphi ]$ has a special property. 
Recall that in the large Hilbert space, one can decompose the string field $\Phi _{-g,p}$ as (\ref{decomposition}), in which a spacetime field $\phi _{g,p}^{\, r \, \eta }$ is multiplied by an $\eta$-exact world-sheet basis. 
Then, for any pairs of $(g,p)$, we find the following relation 
\begin{align}
\label{kernel}
\frac{\partial }{\partial \phi _{g,p}^{\, r \, \eta } } S_{\textsf{bv}} [ \varphi ] = \La \eta \, \xi \, \cZ _{-g,p}^{\, r} , \, \bM \frac{1}{1- \eta \, \varphi } 
\Ra = 0 \, .
\end{align}
This kind of property plays a crucial role in the constrained BV approach. 

\vspace{1mm}

In the rest, we often omit the $r$-label and use the following short notation for brevity: 
\begin{align}
\label{half}
\big{|} \cZ _{-g,p}^{\,\, \eta } \ra \equiv \eta \, \xi \, \big{|} \cZ _{-g,p} \ra \, , 
\hspace{5mm} 
\big{|} \cZ _{-g,p}^{\,\, \xi } \ra \equiv \xi \, \eta \, \big{|} \cZ _{-g,p} \ra \, . 
\end{align} 
Likewise, we often use $| \cZ _{-g,p}^{\, \ast \, \eta } \rangle \equiv \eta \, \xi \, | \cZ _{-g,p}^{\, \ast } \rangle $ and $| \cZ _{-g,p}^{\, \ast \, \xi } \rangle \equiv \xi \, \eta \, | \cZ _{-g,p}^{\, \ast } \rangle $ as (\ref{half}). 
See appendix A for their properties. 
We write $\Phi _{-g,p} = \phi _{g,p}^{\, \eta } \, | \cZ _{-g,p}^{\,\, \eta } \rangle + \phi _{g,p}^{\, \xi } \, | \cZ _{-g,p}^{\,\, \xi } \rangle $ for the decomposition (\ref{decomposition}) of the string field $\Phi _{-g,p}$. 
Then, one can expand the antibracket (\ref{antibracket}) using 
\begin{align*}
\frac{\overset{\leftarrow }{\partial }}{\partial \phi _{g,p} } 
\frac{\overset{\rightarrow }{\partial }}{\partial (\phi _{g,p} )^{\ast } } 
= \frac{\overset{\leftarrow }{\partial }}{\partial \phi _{g,p}^{\,\, \xi } } 
\frac{\overset{\rightarrow }{\partial }}{\partial (\phi _{g,p}^{\,\, \xi })^{\ast } } 
+ \frac{\overset{\leftarrow }{\partial }}{\partial \phi _{g,p}^{\,\, \eta } }
\frac{\overset{\rightarrow }{\partial }}{\partial (\phi _{g,p}^{\,\, \eta } )^{\ast } } \, .
\end{align*}

\subsection{Preliminary: Constrained BV for partially gauge-fixed theory} 

We write $(\varphi )^{\ast }$ for the sum of all string antifields. 
Utilizing $\varphi $ of (\ref{all string fields}) and corresponding $(\varphi )^{\ast }$, we impose the following constraint on the space of string fields 
\begin{subequations} 
\begin{align}
\label{total constraint}
\widehat{\Gamma } \equiv ( \varphi )^{\ast } - \eta \, \varphi \, . 
\end{align}
This $\widehat{\Gamma }$ provides constraint equations on each string field, which introduce the spacetime antifields $(\phi _{g,p}^{r} )^{\ast }$ into the constrained master action $S_{\textsf{bv}} [\varphi ] |_{\widehat{\Gamma }}$\,. 
One can decompose this constraint $\widehat{\Gamma }$ with respect to its spacetime ghost number; for $g \geq 0$, we find 
\begin{align}
\label{total constraint b}
\Gamma _{g,p} & \equiv \frac{1}{\sqrt{2} } \Big[ (\Phi _{1+g,-p} )^{\ast } - \eta \, \Phi _{-g,p} \Big] \, , 
\hspace{5mm} 
\Gamma _{-1-g,-p} \equiv \frac{1}{\sqrt{2} } \Big[ ( \Phi _{-g,p} )^{\ast } - \eta \, \Phi _{1+g,-p} \Big] \, . 
\end{align}
\end{subequations} 
Note that the $g$-label of $\Gamma $ denotes its spacetime ghost number, and these give equations for spacetime fields--antifields on a fixed world-sheet basis. 
The $\xi$-exact components $(\Phi ^{\ast })^{\xi }$ of string antifields $\Phi ^{\ast}= (\Phi ^{\ast })^{\eta } + (\Phi ^{\ast })^{\xi }$ give the first class constraints. 
The second class constraints are imposed between the $\xi$-exact components $\Phi ^{\xi }$ of string fields $\Phi = \Phi ^{\eta } +\Phi ^{\xi }$ and the $\eta $-exact components $(\Phi ^{\ast } )^{\eta } = ( \Phi ^{\xi } )^{\ast }$ of string antifields $\Phi ^{\ast } = (\Phi ^{\ast })^{\eta } + (\Phi ^{\ast })^{\xi }$. 
Since our master action (\ref{simplest bv}) is invariant under these first class $\Gamma $, we focus on the second class $\Gamma $. 
Independent second class constraints give the following nonzero antibracket 
\begin{align}
\label{simplest}
\big{(} \, \Gamma _{g,p}^{(1)} \, , \, \Gamma _{g',p'}^{(2)} \, \big{)} & = 
\delta _{g+g',-1} \, \delta _{p+p',0} \, (-)^{g+1} 
\Big( \eta \, \big{|} \cZ _{-g,p}^{\,\, \xi} \ra \Big)^{(1)} \big{|} \cZ _{-g,p}^{\, \ast \, \eta } \ra ^{(2)} \, . 
\end{align}
We used the relation (\ref{half basis}) and short notations $| \cZ _{-g,p}^{\, \ast \, \eta } \rangle \equiv \eta \, \xi \, | \cZ _{-g,p}^{\, \ast } \rangle $ and $| \cZ _{-g,p}^{\, \ast \, \xi } \rangle \equiv \xi \, \eta \, | \cZ _{-g,p}^{\, \ast } \rangle $ introduced in (\ref{half}).  
We quickly find that the following matrix 
\begin{align} 
\label{inverse of simplest} 
\big{(} \, \Gamma _{g,p}^{(1)}  \, , \, \Gamma _{g',p'}^{(2)} \, \big{)}^{-1} & = 
- \delta _{g+g',-1} \, \delta _{p+p',0} \, 
\big{|} \cZ _{-g',p'}^{\,\, \xi } \ra ^{(1)} \Big{(} \xi \, \big{|} \cZ _{-g',p'}^{\, \ast \, \eta } \ra \Big{)}^{(2)} \,
\end{align} 
gives the inverse of (\ref{simplest}) in the sense of 
\begin{align}
\label{def of inv} 
\sum_{h,q} \sum_{h',q'} 
\big{(} \, \Gamma _{g,p}^{(1)} \, , \, \Gamma _{h,q}^{(1')} \, \big{)}
\cdot 
\big{(} \, \Gamma _{h,q}^{(1')} \, , \, \Gamma _{h',q'}^{(2')} \, \big{)}^{-1}
\cdot 
\big{(} \, \Gamma _{h',q'}^{(2')} \, , \, \Gamma _{g',p'}^{(2)} \, \big{)}
=
\big{(} \, \Gamma _{g,p}^{(1)} \, , \, \Gamma _{g',p'}^{(2)} \, \big{)} \, , 
\end{align}
where the dot denotes the inner product of two states: $| A \rangle ^{(1)} \cdot | B \rangle ^{(1)} = \langle A,B \rangle = \langle A | ^{(1)} \cdot \langle B |^{(1)}$ and $| A \rangle ^{(1)} \cdot | B \rangle ^{(2)} = \langle A | ^{(1)} \cdot \langle B |^{(2)} = 0$\,. 
In particular, in the inner product of (\ref{def of inv}), the matrix (\ref{inverse of simplest}) works as a projector onto fixed world-sheet ghost and picture numbers, and switches the label of the state space. 
By taking inner products on both sides of (\ref{inverse of simplest}), we find 
\begin{subequations} 
\begin{align}
\label{rel}
(-)^{g} \, \big{|} \cZ _{-g,p}^{\, \ast \, \eta } \ra ^{(1)} \cdot 
\Big( \, \Gamma _{-1-g,-p}^{(1)} \, , \, \Gamma _{g',p'}^{(2)} \, \Big) ^{-1} 
\cdot \big{|} \cZ _{1+g',-p'}^{\, \ast \, \eta } \ra ^{(2)} 
= - \delta _{g,g'} \, \delta _{p,p'} \, . 
\end{align}
Note that the half bases satisfy $\langle \cZ _{g,p}^{\, \ast \, \xi } , \cZ _{h,p}^{\, \eta } \rangle = (-)^{g} \langle \cZ _{g,p}^{\, \ast \, \eta } , \cZ _{h,q}^{\, \xi } \rangle = - \delta _{g,h} \, \delta _{p,q}$\,; see appendix A. 
The antibrackets of these second class $\Gamma $ and $S_{\textsf{bv}} [\varphi ]$ are 
\begin{align}
\label{rel b}
\big{(} \, \Gamma _{g,p} \, , \, S_{\textsf{bv}} [\varphi ] \, \big{)} 
& = - \Gamma _{g,p} \frac{\overset{\leftarrow }{\partial } }{\partial (\phi ^{\xi })^{\ast } } \, \frac{\overset{\rightarrow }{\partial } }{\partial \phi ^{\xi }} S_{\textsf{bv}} [\varphi ] 
= - \big{|} \cZ _{1+g,-p}^{\, \ast \, \eta } \ra \, \La \frac{\partial \varphi }{\partial \phi _{-1-g,-p}^{\,\, \xi } } , \, \bM \frac{1}{1-\eta \, \varphi } \Ra 
\no 
& = - \big{|} \cZ _{1+g,-p}^{\, \ast \, \eta } \ra \, \la \cZ _{1+g,-p}^{\,\, \xi } \big{|} \, \eta \, \xi \, \big{|} \, \bM \frac{1}{1-\eta \, \varphi } \ra \, , 
\\ 
\big{(} \, S_{\textsf{bv}} [\varphi ] \, , \Gamma _{-1-g,-p} \, \big{)} 
& = S_{\textsf{bv}} [\varphi ] \frac{\overset{\leftarrow }{\partial } }{\partial \phi ^{\xi } } \, \frac{\overset{\rightarrow }{\partial } }{\partial (\phi ^{\xi })^{\ast }} \Gamma _{g,p}
= (-)^{g} \La \bM \frac{1}{1-\eta \, \varphi } , \, \frac{\partial \varphi }{\partial \phi _{g,p}^{\,\, \xi } } \Ra \, 
\big{|} \cZ _{-g,p}^{\, \ast \, \eta } \ra 
\no 
& = (-)^{g} \la \bM \frac{1}{1-\eta \, \varphi } \big{|} \cZ _{-g,p}^{\,\, \xi } \ra \, \big{|} \cZ _{-g,p}^{\, \ast \, \eta } \ra \, . 
\end{align}
\end{subequations} 
Note that $- \langle \cZ _{1+g,-p}^{\,\, \xi } | \, \eta = (-)^{g} \langle \, \eta \, \cZ _{1+g,-p}^{\,\, \xi } | = (-)^{gp} \langle \cZ _{2+g,-p-1}^{\,\, \eta } | = \langle \cZ _{-g,p}^{\, \ast \, \eta } |$ holds. 
Using these relations (\ref{rel}-c), the definition of dual bases (\ref{dual BPZ basis}), and defining properties (\ref{half basis}), we obtain 
\begin{align}
\label{constrained master eq}
\big{(} \, S_{\textsf{bv}} \, , \, S_{\textsf{bv}} \, \big{)}_{\Gamma } 
& \equiv \big{(} \, S_{\textsf{bv}} \, , \, S_{\textsf{bv}} \, \big{)} 
- \sum_{g,p} \sum_{g' , p' } \big{(} \, S_{\textsf{bv}} \, , \, \Gamma _{g,p} \, \big{)} \cdot 
\big{(} \, \Gamma _{g,p} \, , \, \Gamma _{g',p'} \, \big{)}^{-1} \cdot 
\big{(} \, \Gamma _{g',p'} \, , \, S_{\textsf{bv}} \, \big{)}  
\no
& = \sum_{g,p} \sum_{g',p'} 
\la \bM \frac{1}{1-\eta \, \varphi } \big{|} \cZ _{-g,p}^{\,\, \xi } \ra \, 
\delta _{g,g'} \, \delta _{p,p'}
\Big{[} - \la \cZ _{1+g',-p'}^{\,\, \xi } \big{|} \, \eta \, \Big{]} \, \xi \, \big{|} \bM \frac{1}{1-\eta \, \varphi } \ra 
\no
& = - \La \bM \frac{1}{1-\eta \, \varphi } , \, \xi \, \bM \frac{1}{1-\eta \, \varphi } \Ra = 0 \, . 
\end{align} 
The action $S_{\textsf{bv}}$ and constraint $\widehat{\Gamma }$ give a solution of the constrained BV master equation. 
Here, we used the mutual commutativity $\ld \Eta , \, \bM \rd = 0$ and the cyclic $A_{\infty }$ relation of $\bM$, 
\begin{align*}
\La \bM \frac{1}{1-\eta \, \varphi } , \, \xi \, \bM \frac{1}{1-\eta \, \varphi } \Ra 
& = \sum_{k=1}^{\infty } \sum_{l=1}^{\infty } \Big[ \frac{k}{k+l} + \frac{l}{k+l} \Big] 
\La \bM_{k} \frac{1}{1-\eta \, \varphi } , \, \xi \, \bM_{l} \frac{1}{1-\eta \, \varphi } \Ra 
\no 
& = \sum_{k=1}^{\infty } \sum_{l=1}^{\infty } \frac{2}{k+l} 
\La \, \eta \, \varphi , \, 
\bM_{k} \frac{1}{1-\eta \, \varphi } \otimes \xi \Big[ \pi _{1} \, \bM_{l}  \frac{1}{1-\eta \, \varphi } \Big] \otimes 
\frac{1}{1-\eta \, \varphi } \Ra \, 
\no 
& = \sum_{n=1}^{\infty }  \frac{1}{n+1} \sum_{m=0}^{n-1} 
\La \, \varphi , \, \Ld \, \bM_{m+1} , \, \bM_{n-m} \, \Rd  \frac{1}{1-\eta \, \varphi } \Ra \, . 
\end{align*}

Note however that, as can be seen in \cite{Berkovits:2012np}, the constrained BV approach based on (\ref{extra ghost string field}), (\ref{Berkovits BV}), and (\ref{total constraint}) works well just for a \textit{partially gauge-fixed} theory. 
Although it indeed gives a solution of the constrained master equation as (\ref{constrained master eq}), we find that $S_{\textsf{bv}} [ \varphi ] |_{\Gamma }$ has an undesirable property as a proper master action for the \textit{large} theory: 
$\Phi _{-g,p=g}$ for $g>0$ behaves as a nontrivial auxiliary ghost field. 
We write $(\varphi _{-} )^{\ast }$ for the sum of the (original) string antifields $\{ (\Phi _{-g,p} )^{\ast } | g \geq 0 , \, 0 \leq p \leq g \}$ corresponding to the (original) string fields $\{ \Phi _{-g,p} \, | \, g\geq 0, \, 0 \leq p \leq g \} $. 
On the constraint surface defined by (\ref{total constraint}), the above $S_{\textsf{bv}} [\varphi ]$ takes the following form 
\begin{align}
\label{simplest bv}
S_{\textsf{bv} } [\varphi ] |_{\Gamma } & = \int _{0}^{1} dt \, \la \xi \, 
( \eta \, \varphi _{-} + (\varphi _{-})^{\ast } ), \, \bM \frac{1}{1 - t \, ( \eta \, \varphi _{-} - (\varphi _{-} )^{\ast } ) } \ra 
\no 
& = \frac{1}{2} \la \Phi  , \, Q \, \eta \, \Phi  \ra 
+ \sum_{g\geq 0} \sum_{p=0}^{g} \la (\Phi _{-g,p})^{\ast } , \, Q \, \Phi _{-1-g,p} \ra 
\no & \hspace{20mm} 
+ \sum_{g\geq 0} \sum_{p=0}^{g} 
\la (\Phi _{-g,p} )^{\ast } , \, \sum_{n>1} \bM _{n} \frac{1}{1- \eta \, \varphi _{-} - ( \varphi _{-} )^{\ast } } \ra \, . 
\end{align}
The ghost string field $\Phi _{-g,p=g}$ for $g>0$ does not have its kinetic term. 
This line of ghosts does not exist in a partially gauge-fixed theory $S[\Phi ^{\xi } ]$ from the beginning, and then it gives a correct proper master action. 
By contrast, the large theory $S[\Phi ]$ requires this line to describe the gauge invariance generated by counting $\Phi ^{\eta }$ of $S[\Phi = \Phi ^{\xi } + \Phi ^{\eta }]$ in spite of $S[\Phi ] = S[\Phi ^{\xi } ]$. 

\vspace{1mm}

We would like to emphasise that this kind of problem (or ambiguity) occurs in every superstring field theories based on the large Hilbert space, even for the conventional BV approach, when we focus on the \textit{spacetime} fields or fine parts $\varphi _{a}$ of string fields $\varphi = \varphi _{1} + \cdots + \varphi _{n}$. 
Let us consider the kinetic term of (\ref{S}). 
Although string fields live in the large Hilbert space, the kinetic term $K[\Phi ]$ satisfies the property (\ref{kernel}), namely, $K[\Phi ] = K[\Phi ^{\xi } ]$ holds for $\Phi = \Phi ^{\xi } + \Phi ^{\eta }$. 
It implies that the $(\phi _{0,0}^{\eta })^{\ast }$-dependence becomes irrelevant in the master equation 
\begin{align*}
\bigg[ K_{\textsf{bv}} \frac{\overset{\leftarrow }{\partial }}{\partial (\phi _{0,0}^{\eta })^{\ast }} \bigg] \, 
 \frac{\overset{\rightarrow }{\partial }}{\partial \phi _{0,0}^{\eta }} K_{\textsf{bv}} = 0 \, . 
\end{align*}
The master action $K_{\textsf{bv}}$ can be any functional of spacetime antifields $(\phi _{0,0}^{\eta })^{\ast }$; for example, one could choose $\langle (\Phi ^{\eta })^{\ast } , \eta \, \Phi _{-1,1} \rangle = 0$\,. 
Then, the kinetic term of the master action (\ref{free}), which was proposed in \cite{Kroyter:2012ni} on the basis of the conventional BV approach, reduces to 
\begin{align} 
\label{partially gauge-fixed free}
K_{\textsf{bv}} = \frac{1}{2} \la \Phi ^{\xi } 
, \, Q \, \eta \, \Phi ^{\xi } 
\ra 
+  \la \Phi ^{\ast } , \, Q \, \Phi _{-1,0} \ra 
+ \sum_{g \geq 1} \sum_{p=0}^{g-1} \la ( \Phi _{-g,p} )^{\ast } , \, Q \, \Phi _{-1-g,p} + \eta \, \Phi _{-1-g,p+1} \ra \, , 
\end{align}
where the string antifield is defined by (\ref{naive string antifield}) and $\langle (\Phi ^{\xi })^{\ast } , \eta \, \Phi _{-1,1} \rangle = 0$\,. 
This is nothing but the kinetic term of a proper BV master action for the partially gauge-fixed theory $K = K[\Phi ^{\xi }]$.
As (\ref{partially gauge-fixed free}) is proper for the partially gauge-fixed theory $K=K[\Phi ^{\xi }]$ but not appropriate for the large theory $K=K[\Phi ]$, the above pair $(S_{\textsf{bv}} , \varphi , \widehat{\Gamma } )$ just gives a correct proper BV master action for \textit{partially gauge-fixed} theory in the large Hilbert space. 

\vspace{1mm}

We would like to construct a constrained BV master action for the \textit{large} theory, which we explain in the rest of this section. 
We investigate appropriate pairs of $(S_{\textsf{bv}} , \varphi , \widehat{\Gamma } )$ which give kinetic terms for these $\Phi _{-g,p=g}$ on the basis of three different approaches. 

\subsection{Constrained BV master action based on improved constraints}

We show that the kinetic terms of $S_{\textsf{bv}} [\varphi ]|_{\Gamma }$ in (\ref{simplest bv}) can be remedied by improving constraint equations, keeping the form of $S_{\textsf{bv}} [\varphi ]$ and the assembly (\ref{extra ghost string field}) of extra string fields.
The key property is (\ref{kernel}). 
Note that for $0\leq p \leq g$, the above constraints can be written as 
\begin{subequations} 
\begin{align} 
\label{improved constraint a}
\Gamma _{g,p} & \equiv \frac{1}{\sqrt{2}} \Big[ (\Phi _{1+g,-p}^{\xi } )^{\ast } - \eta \, \Phi _{-g,p} \Big] \, , 
\hspace{5mm} 
\Gamma _{-1-g,-p}  \equiv \frac{1}{\sqrt{2}} \Big[ (\Phi _{-g,p}^{\xi } )^{\ast } - \eta \, \Phi _{1+g,-p} \Big] \, , 
\end{align}
in which $(\Phi ^{\eta })^{\ast }$ does not appear. 
In addition to these, for $0 \leq p \leq g$, we impose the following (nonlinear) constraint equations\footnote{These constraints are weaker than the following type of linear constraint, 
\begin{align*} 
\widetilde{\gamma }_{-1-g,-p} & \equiv \big[ \delta (\phi _{g,p}^{\, \eta } )^{\ast } - \delta \phi _{-1-g,-1-p} \big] \, \bM _{\eta } \, \big{|} \cZ _{1+g,-1-p} \ra \, , 
\hspace{3mm}  
\bM _{\eta } \, | \cZ _{g,p} \rangle \equiv \big[ \pi _{1} \, \bM \frac{1}{1-\eta \, \varphi } \otimes | \cZ _{g,p} \rangle  \otimes \frac{1}{1-\eta \, \varphi } \big]_{g,p} \, .
\end{align*} 
Roughly, the constraints (\ref{improved constraint}) are written as $\delta \gamma _{-1-g,-p} = \big[ \delta (\phi _{g,p}^{\,\, \eta } )^{\ast } - \sum _{h,q} \delta \phi _{h,q}^{\,\, \eta } A_{g-h,p-q}^{(2+g,-1-p)} \big] | \cZ _{2+g,-1-p}^{\,\, \xi } \rangle $, where $A_{s}^{(g,p)}$ denotes that of $\bM _{\eta } A = \sum_{s,g,p} A_{s}^{(g,p)} | \cZ _{g,p} \rangle $ and has spacetime ghost number $s$. 
} 
\begin{align} 
\label{improved constraint} 
\gamma _{g,p}  \equiv (\Phi _{1+g,-p}^{\,\, \eta } )^{\ast } - \bigg[ \xi \, \bM \frac{1}{1- \eta \, \varphi } \bigg] _{1-g,p-1} \, , 
\hspace{5mm}  
\gamma _{-1-g,-p}  \equiv (\Phi _{g,p}^{\,\, \eta } )^{\ast } - \bigg[ \xi \, \bM \frac{1}{1- \eta \, \varphi }  \bigg{]}_{2+g,-1-p} \, , 
\end{align}
\end{subequations} 
where $[A]_{g,p}$ denotes the components of $A$ which have world-sheet ghost number $g$ and picture number $p$\,. 
For example, $[\varphi ]_{-g,p} = \Phi_{-g,p}$ for (\ref{all string fields}). 
As will see, the above constraints $\gamma $ provides the kinetic terms $\langle (\Phi _{-g,p}^{\,\, \eta })^{\ast } , \, \eta \, \Phi _{-g,p}^{\,\, \xi } \rangle $ in $S_{\textsf{bv}}[\varphi ] |_{\Gamma , \gamma }$. 
Since these constraints are second class, we have to consider the Dirac antibracket based on $\Gamma $ and $\gamma $\,: 
\begin{align*} 
\big{(} \, S_{\textsf{bv}}  , \, S_{\textsf{bv}} \, \big{)}_{\Gamma , \gamma } & \equiv 
\big{(} \, S_{\textsf{bv}} , \, S_{\textsf{bv}} \, \big{)} - 
\big{(} \, S_{\textsf{bv}} , \, \Gamma \, \big{)} \! \cdot \! 
\big{(} \, \Gamma \, , \, \Gamma ' \, \big{)}^{-1} \! \! \cdot 
\big{(} \, \Gamma ' \, , \, S_{\textsf{bv}} \big{)}
 - 
\big{(} \, S_{\textsf{bv}} , \, \gamma \, \big{)} \! \cdot \! 
\big{(} \, \gamma \, , \, \Gamma \, \big{)}^{-1} \! \! \cdot  
\big{(} \, \Gamma \, , \, S_{\textsf{bv}} \big{)}
\no & \hspace{10mm} 
- 
\big{(} \, S_{\textsf{bv}} , \, \Gamma \, \big{)} \! \cdot \!
\big{(} \, \Gamma \, , \, \gamma \, \big{)}^{-1} \! \! \cdot 
\big{(} \, \gamma \, , \, S_{\textsf{bv}} \big{)}
- 
\big{(} \, S_{\textsf{bv}}  \, , \, \gamma \, \big{)} \! \cdot \! 
\big{(} \, \gamma , \, \gamma ' \, \big{)}^{-1} \! \! \cdot  
\big{(} \, \gamma ' \, , \, S_{\textsf{bv}} \big{)} \, . 
\end{align*} 
The inverse matrices labeled by $\gamma $ take complicated forms because of $\gamma $'s nonlinearity. 
However, by construction of $S_{\textsf{bv}} [\varphi ]$, we do not have to know the explicit form of $(\gamma , \Gamma )^{-1}$, $(\Gamma , \gamma )^{-1}$, or $(\gamma , \gamma ')^{-1}$ to solve the master equation. 
Because of $\frac{\partial }{\partial \phi ^{\ast }} S_{\textsf{bv}} [\varphi ] = \frac{\partial }{\partial \phi ^{\eta }} S_{\textsf{bv}} [\varphi ] =0$ and $\frac{\partial }{\partial (\phi ^{\xi })^{\ast }} \gamma = 0$, we find  
\begin{align}
\label{invariant constraint}
\big{(} \, \gamma _{g,p} \, , \, S_{\textsf{bv}} [\varphi ] \, \big{)} 
& = -  \gamma _{g,p} \frac{\overset{\leftarrow }{\partial }}{\partial (\phi ^{\eta })^{\ast }} \frac{\overset{\rightarrow }{\partial }}{\partial \phi ^{\eta } } S_{\textsf{bv}} [\varphi ] 
= 0  \, , 
\end{align}
and thus the constrained master action (\ref{Berkovits BV}) is invariant\footnote{As another option, one could introduce $\gamma $ as the first class constraints which preserve $S_{\textsf{bv}} [\varphi ]$.} under new constraints $\gamma$. 
Hence, these new constraints $\gamma $ give no contribution to $(S_{\textsf{bv}} , S_{\textsf{bv}} )_{\Gamma , \gamma }$ and we find 
\begin{align*}
\big{(} \, S_{\textsf{bv}}  \, , \, S_{\textsf{bv}} \, \big{)}_{\Gamma , \gamma } 
=  - \big{(} \, S_{\textsf{bv}}  \, , \, \Gamma \, \big{)} 
\big{(} \, \Gamma  \, , \, \Gamma ' \, \big{)}^{-1} 
\big{(} \, \Gamma ' \, , \, S_{\textsf{bv}} \, \big{)}
= \big{(} \, S_{\textsf{bv}}  \, , \, S_{\textsf{bv}} \, \big{)}_{\Gamma } = 0 \, . 
\end{align*}
On the constrained subspace, we can rewrite the master action as follows, 
\begin{align*}
S_{\textsf{bv}} [\varphi ] |_{\Gamma , \gamma } & = \int_{0}^{1} dt \, \La \xi \, ( \eta \, \varphi _{-}^{\,\, \xi } + \eta \, \varphi _{\textsf{ex}}^{\,\, \xi } ) , \, \bM \frac{1}{1- t \, \eta \, \varphi }  \Ra 
\no 
& = \sum_{g\geq 0} \sum_{p=0}^{g} \bigg[ 
\La \eta \, \Phi _{-g,p}^{\,\, \xi } , \, \xi \, \bM \frac{1}{1- t \, \eta \, \varphi  }  \Ra 
+ \La \eta \, \Phi _{1+g,-p}^{\,\, \xi } , \, \bxi \, \bM \frac{1}{1- t \, \eta \, \varphi _{-} - t \, \eta \, \varphi _{\textsf{ex}} }  \Ra 
\bigg] 
\no 
& = \sum_{g \geq 0} \sum_{p=0}^{g} \La (\Phi _{-g,p}^{\,\, \eta } )^{\ast } , \, \Eta \, \Phi _{-g,p}^{\,\, \xi } \Ra 
+ \sum_{g \geq 0} \sum_{p=0}^{g} \La ( \Phi _{-g,p}^{\,\, \xi } )^{\ast } , \, \bxi \, \bM \frac{1}{1- t \, \eta \, \varphi _{-} - t \, ( \varphi _{-}^{\,\, \xi } )^{\ast } } \Ra \, . 
\end{align*}
In contrast to (\ref{simplest bv}), it includes string antifields for $\Phi _{-g,p}^{\,\, \eta }$ and has the kinetic terms for all original string fields and their string antifields; it is proper for the \textit{large} theory. 

\vspace{1mm} 

In the constrained BV approach, the BV transformation of $\varphi $ is given by $\delta _{\textsf{bv}} \varphi = ( \varphi , S_{\textsf{bv}} )_{\Gamma ,\gamma }$\,. 
By construction of the constraints, $\delta _{\textsf{bv}} \varphi $ has an orthogonally decomposed form. 
Note that while $( \varphi ^{\xi } , \Gamma )$ is the antibracket of $\xi$-exact $\varphi ^{\xi }$ and $\eta $-exact $\Gamma $, $(\varphi ^{\eta } , \gamma )$ is the antibracket of $\eta $-exact $\varphi ^{\eta }$ and $\xi$-exact $\gamma $\,. 
We set $\Omega \equiv \xi \, ( \varphi ^{\eta } , \, \gamma \, ) \cdot ( \gamma , \, \Gamma )^{-1} \cdot ( \Gamma , \, S_{\textsf{bv}} )$ using the unknown inverse. 
We find that the BV transformation $\delta _{\textsf{bv}}\varphi = \delta _{\textsf{bv}} \varphi ^{\xi } + \delta _{\textsf{bv}} \varphi ^{\eta }$ is given by 
\begin{align*}
\delta _{\textsf{bv}} \varphi ^{\xi } = \big{(} \, \varphi ^{\xi } \, , \, S_{\textsf{bv}} \, \big{)}_{\Gamma , \gamma } = \pi _{1} \, \bxi \, \bM \frac{1}{1-\eta \, \varphi } \, , 
\hspace{5mm} 
\delta _{\textsf{bv}} \varphi ^{\eta } = \big{(} \, \varphi ^{\eta } \, , \, S_{\textsf{bv}} \, \big{)}_{\Gamma , \gamma } = \Eta \, \Omega \, . 
\end{align*}

\subsection{Alternative: Modifying extra string fields}

There is another option to obtain a proper BV master action for the large theory. 
Let us consider the extra string fields of (\ref{extra ghost string field}). 
We decompose them as follows 
\begin{subequations} 
\begin{align}
\label{modified extra}
\Phi _{1+g,-p} \equiv \phi _{-1-g,-p}^{\,\, \eta } \, \big{|} \cZ _{1+g,-p}^{\,\, \eta } \ra + \phi _{-1-g,-p}^{\,\, \xi } \, \big{|} \cZ _{1+g,-p}^{\,\, \xi } \ra \,\, . 
\end{align} 
For these (\ref{modified extra}), we introduce additional extra string fields $\bar{\Phi }$ assembled as 
\begin{align} 
\bar{\Phi }_{1+g,-1-p} \equiv \phi _{-1-g,-p}^{\,\, \xi } \, \big{|} \cZ _{1+g,-1-p}^{\,\, \eta } \ra + \phi _{-1-g,-p}^{\,\, \eta } \, \big{|} \cZ _{1+g,-1-p}^{\,\, \xi } \ra \, .  
\end{align} 
\end{subequations} 
While the set of extra spacetime ghost fields is not changed, the assemblies of extra ghost string fields are modified by considering additional world-sheet bases. 
For simplicity, we consider the sum of these extra ghost string fields as follows 
\begin{align*} 
\varphi '_{\textsf{ex}} \equiv \sum_{g = 0}^{\infty } 
\sum_{p=0}^{g+1} \Phi _{1+g,-p}' \, , 
\hspace{5mm} 
\Phi '_{1+g,-p} \equiv ( 1 - \delta _{p,g+1} ) \, \Phi _{1+g,-p} + (1 - \delta _{p,0} ) \, \bar{\Phi }_{1+g,-p} \, . 
\end{align*} 
Note that the $p$-label of $\Phi ^{\prime }_{1+g,-p}$ runs from $0$ to $g+1$ unlike that of $\Phi _{1+g,-p}$\,. 
Now, the sum of all string fields (\ref{all string fields}) is replaced by $\varphi ' \equiv \varphi _{-} + \varphi '_{\textsf{ex}}$, and the constrained BV master action $S_{\textsf{bv}} = S_{\textsf{bv}} [\varphi ']$ is a functional of $\varphi '$\,. 
Namely, $S_{\textsf{bv}}$ is changed via modifying extra ghost string fields from $\varphi $ to $\varphi '$\,. 
Since this $S_{\textsf{bv}}[\varphi ']$ includes kinetic terms for all string fields as follows 
\begin{align*}
S_{\textsf{bv}} [\varphi '] = \frac{1}{2} \la \, \Phi , \, Q \, \eta \, \Phi \, \ra 
+ \sum_{g=0}^{\infty } \sum_{p=0}^{g+1} \la \, \Phi _{1+g,-p}' , \, Q \, \eta \, \Phi _{-1-g,p} \, \ra 
+ \cdots \, ,  
\end{align*} 
the resultant $S_{\textsf{bv}} [ \varphi ' ] |_{\Gamma '}$ has corresponding terms under appropriate constraints $\Gamma '$\,. 
Instead of (\ref{total constraint}), for instance, we can impose the same type of \textit{second} class constraint 
\begin{align*}
\widehat{\Gamma }' \equiv (\varphi ')^{\ast } - \eta \, \varphi ' \, , 
\end{align*}
where $(\varphi ' )^{\ast }$ denotes the sum of (the $\eta $-exact parts of) all string antifields 
\begin{align*}
( \varphi ' )^{\ast } = \bigg{[} \Phi ^{\ast } + \sum_{g > 0} \sum_{p=0}^{g} (\Phi _{-g,p})^{\ast } + \sum_{g \geq 0} \sum_{p=0}^{g+1} ( \Phi _{1+g,p}^{\prime } )^{\ast } \bigg{]}^{\eta } \, . 
\end{align*} 
The above $\widehat{\Gamma }^{\prime }$ provides the same type of second class constraint as (\ref{total constraint b}) for $0 \leq p \leq g$. 
By contrast, for $p = g+1$, we set $\Phi _{-g,g+1} = (\Phi _{-g,g+1})^{\ast } =0$ and constraints reduce to simple ones: 
\begin{align*}
\Gamma _{g,p}^{\prime } = \frac{1}{\sqrt{2}} \Big[ (\Phi _{1+g,-p}^{\prime })^{\ast } -  \underbrace{\eta \, \Phi _{-g,p} }_{\underset{(g<p)}{\longrightarrow } 0} \Big] \, , \hspace{5mm} 
\Gamma _{-g-1,-p}^{\prime } = \frac{1}{\sqrt{2}} \Big[ \underbrace{(\Phi _{-g,p})^{\ast }}_{\underset{(g<p)}{\longrightarrow } 0} - \eta \, \Phi _{1+g,-p}^{\prime } \Big] \, . 
\end{align*}
These constraints assign (physical) antifields to world-sheet bases labeled by different picture numbers and the constrained string antifields are given by (\ref{example}). 
In this case, the string field representation of the antibracket does not have the Darboux form. 
Note however that the spacetime antibracket itself can always take the Darboux form and thus as we will see, the constrained master equation holds in the same manner as (\ref{constrained master eq}). 

\vspace{1mm} 

Note that we consider the second class and do not impose the first class constraints on the extra string fields this time. 
While the $\eta$-acted extra string field is given by 
\begin{subequations}
\begin{align}
\label{ex' a}
\eta \, \Phi _{1+g,-p}^{\prime } = \Big{[} \phi _{-1-g,-p}^{\, \xi } + \phi _{-1-g,1-p}^{\, \eta } \Big{]} (-)^{1+g} \eta \, \big{|} \cZ _{1+g,-p}^{\, \xi } \ra \, , 
\end{align}
the string antifield for the extra string field $\Phi _{1+g,-p}^{\prime }$ is given by 
\begin{align}
\label{ex' b}
\big{(} \Phi _{1+g,-p}^{\prime } \big{)}^{\ast } = \Big[ (\phi _{-1-g,-p}^{\, \xi })^{\ast } + (\phi _{-1-g,1-p}^{\, \eta })^{\ast } \Big] \big{|} \cZ _{1+g,-p}^{\, \ast \, \eta } \ra \, . 
\end{align}
\end{subequations} 
The second class constraints are imposed on these $\eta $-exact states (\ref{ex' a}-b). 
Let us check that the pair $(S_{\textsf{bv}} [ \varphi ' ] , \, \Gamma ' )$ solves the constrained BV master equation. 
We consider the antibracket of constraints, whose $\phi ^{\xi }$-part is the same as (\ref{simplest}). 
We find 
\begin{align}
\label{2'=2+1}
\big{(} \, \Gamma _{g,p}^{\prime \, (1)} \, , \, \Gamma _{g',p'}^{\prime \, (2)} \, \big{)} = 
\big{(} \, \Gamma _{g,p}^{(1)} \, , \, \Gamma _{g',p'}^{(2)} \, \big{)} 
+ \Gamma _{g,p}^{\prime \, (1)} \bigg[ 
\frac{\overset{\leftarrow }{\partial }}{\partial \phi ^{\eta } } \frac{\overset{\rightarrow }{\partial }}{\partial (\phi ^{\eta })^{\ast }} 
- 
\frac{\overset{\leftarrow }{\partial }}{\partial (\phi ^{\eta })^{\ast } } \frac{\overset{\rightarrow }{\partial }}{\partial \phi ^{\eta } } \bigg] 
\Gamma _{g',p'}^{\prime \, (2)} \, . 
\end{align} 
In particular, the second term of (\ref{2'=2+1}) vanishes for $p=0$. 
By construction of extra string fields--antifields (\ref{ex' a}-b), one can quickly find that (\ref{2'=2+1}) is given by 
\begin{align*}
\big{(} \, \Gamma _{g,p}^{\prime \, (1)} \, , \, \Gamma _{g',p'}^{\prime \, (2)} \, \big{)} 
= \delta _{g+g',-1} \, \delta _{p+p',0} (-)^{g+1} 
\sum_{k=0,1} \Big{(} \eta \, \big{|} \cZ _{-g,p+k}^{\, \xi } \ra \Big{)}^{(1)} \big{|} \cZ _{-g,p+k}^{\, \ast \, \eta } \ra ^{(2)} \, 
\end{align*}
for $0 < p \leq g$. 
The $k=0$ parts arise from the first part of (\ref{2'=2+1}); the $k=1$ parts arise from the second part of (\ref{2'=2+1}). 
Likewise, the following relation holds for $p=g+1$, 
\begin{align*}
\big{(} \, \Gamma _{g,p}^{\prime \, (1)} \, , \, \Gamma _{g',g'+1}^{\prime \, (2)} \, \big{)} 
= \delta _{g+g',-1} \, \delta _{p+(g'+1),0} (-)^{g+1} \Big{(} \eta \, \big{|} \cZ _{-g,p+1}^{\, \xi } \ra \Big{)}^{(1)} \big{|} \cZ _{-g,p+1}^{\, \ast \, \eta } \ra ^{(2)} \, . 
\end{align*}
In the sense of (\ref{def of inv}), the inverse matrix of (\ref{2'=2+1}) is given by 
\begin{align} 
\label{inv'}
\big{(} \, \Gamma _{g,p}^{\prime \, (1)} \, , \, \Gamma _{g',p'}^{\prime \, (2)} \, \big{)} ^{-1} 
& = \delta _{g+g',-1} \bigg[ 
\delta _{p,0} \big{|} \cZ _{-g',1}^{\, \xi } \ra ^{(1)} \Big{(} \xi \, \big{|} \cZ _{-g',1}^{\, \ast \, \eta } \ra \Big{)}^{(2)} 
+ \delta _{p,g+1} \big{|} \cZ _{-g',g'}^{\, \xi } \ra ^{(1)} \Big{(} \xi \, \big{|} \cZ _{-g',g'}^{\, \ast \, \eta } \ra \Big{)}^{(2)} \bigg] 
\no & \hspace{15mm} 
- \delta _{g+g',-1} \, \delta _{p+p',0} 
\sum_{k=0,1} \big{|} \cZ _{-g',p'+k}^{\, \xi } \ra ^{(1)} \Big{(} \xi \, \big{|} \cZ _{-g',p'+k}^{\, \ast \, \eta } \ra \Big{)}^{(2)} \, . 
\end{align} 
Note that $| \cZ _{g,p+k}^{\, \ast } \rangle \cdot | \cZ_{g,p+l} \rangle = - \delta _{k,l}$ holds. 
The Dirac antibracket is defined by using (\ref{inv'}). 
Let us consider the antibracket of $\Gamma _{g,p}^{\prime }$ and $S_{\textsf{bv}} [\varphi ^{\prime }]$, whose $\phi ^{\xi }$-part is the same as (\ref{rel b}). 
We find 
\begin{align*}
\big{(} \, \Gamma _{g,p}^{\prime } \, , S_{\textsf{bv}} [\varphi ^{\prime }] \, \big{)} 
& = \big{(} \, \Gamma _{g,p} \, , S_{\textsf{bv}} [\varphi ^{\prime }]\, \big{)} 
+ \Gamma _{g,p}^{\prime } 
\bigg[ \frac{\overset{\leftarrow }{\partial }}{\partial \phi ^{\eta } } \frac{\overset{\rightarrow }{\partial }}{\partial (\phi ^{\eta })^{\ast }} 
- \frac{\overset{\leftarrow }{\partial }}{\partial (\phi ^{\eta })^{\ast } } \frac{\overset{\rightarrow }{\partial }}{\partial \phi ^{\eta } } \bigg] 
S_{\textsf{bv}} [\varphi ^{\prime } ]  
\no 
& =
\big{(} \, \Gamma _{g,p} \, , \, S_{\textsf{bv}} [\varphi ] \, \big{)} 
- \big{|} \cZ _{1+g,1-p}^{\, \ast \, \eta } \ra \, \la \cZ _{1+g,1-p}^{\,\, \xi } \big{|} \, \eta \, \xi \, \big{|} \, \bM \frac{1}{1-\eta \, \varphi } \ra \, .  
\end{align*}
The second term is not zero because of $\frac{\partial }{\partial \phi ^{\eta }} S_{\textsf{bv}} [\varphi ^{\prime } ] \not= 0$ unlike (\ref{kernel}) and contracts with the $k=1$ parts of (\ref{inv'}) only in the Dirac antibracket, which gives the same contribution as (\ref{constrained master eq}) after the sum. 
The first term contracts with the $k=0$ parts only and there are no other contractions in the Dirac antibracket. 
Hence, we obtain $(S_{\textsf{bv}} , S_{\textsf{bv}} )_{\Gamma ^{\prime } } = (S_{\textsf{bv}} , S_{\textsf{bv}} )_{\Gamma } + (S_{\textsf{bv}} , S_{\textsf{bv}} )_{\Gamma } = 0$\,. 


\clearpage 

\subsection{Other constrained BV master actions: Switching $\bM$ to $\Eta$} 

Can we simplify the nonlinear constraints (\ref{improved constraint}) by retaking the constrained BV action $S_{\textsf{bv}}$ of (\ref{Berkovits BV})?---it is possible. 
The construction of the Dirac antibracket suggests that such a constrained BV action will be obtained by switching a part of (\ref{Berkovits BV}) which generates nonlinear $\bM$-gauge transformations to the terms which generate linear $\Eta $-gauge transformations as \cite{Matsunaga:2017phm}. 
We introduce the following extra string fields $\Psi _{2+g,-1-p}$ and their sum $\psi $, 
\begin{align}
\label{shifted extra ghost}
\psi \equiv \sum_{g \geq 0} \sum_{p=0}^{g} \Psi _{2+g,-1-p} \, , 
\hspace{5mm} 
\Psi _{2+g,-1-p} \equiv \sum \phi _{-1-g,p} \, \big{|} \cZ _{2+g,-1-p} \ra \, . 
\end{align}
It has the same form as the original string antifield $(\Phi _{-g,p})^{\ast }$. 
Note that the $\eta $-exact component of $\Psi _{2+g,-1-p}$ equals to $\eta \, \Phi _{1+g,-p}$\,, and thus these new extra string fields are Grassmann odd: $(-)^{\varphi } = (-)^{\psi +1}$. 
We split the sum of all string fields (\ref{all string fields}) into two parts 
\begin{align*}
\varphi = \varphi _{1} + \varphi _{2} \, . 
\end{align*}
One can consider any splittings as long as $\varphi _{1}$ include the dynamical field $\Phi $\,. 
As a functional of these $\varphi _{1}$, $\varphi _{2}$ and $\psi $, we consider the following action, 
\begin{align}
\label{switched BV}
S_{\textsf{bv}} [\varphi _{1} ; \varphi _{2} , \psi ] 
= \int_{0}^{1} dt \, \LA \, \varphi _{1} \, , \, \bM \frac{1}{1- t \, \eta \, \varphi _{1} } \, \RA 
+ \la \, \psi \, , \, \Eta \, \varphi _{2} \, \ra \, .
\end{align}
It reduces to the original action (\ref{original action}) if we set all extra fields zero. 
We write $S_{1}$ for the first term and $S_{2}$ for the second term: $S_{\textsf{bv}} [\varphi _{1} ; \varphi _{2} , \psi ]= S_{1}[\varphi _{1}] + S_{2}[\varphi _{2} , \psi ]$\,. 
Note that the $\eta $-exact components of $\psi$ do not appear in the second term. 
The variation of $S_{\textsf{bv}}$ is given by 
\begin{align*}
\delta S_{\textsf{bv}} & = \La \, \delta \varphi _{1} , \, \bM \frac{1}{1 - \eta \, \varphi _{1}} \, \Ra 
+ \la \, \delta \varphi _{2} \, , \, \Eta \, \psi \, \ra 
+ \la \, \delta \psi \, , \, \Eta \, \varphi _{2} \, \ra \, . 
\end{align*} 
We find that the action $S_{\textsf{bv}}=S_{1}[\varphi _{1}] + S_{2}[\varphi _{2} , \psi ]$ is invariant under the gauge transformations 
\begin{align*} 
\delta \varphi _{1} = \pi _{1} \, \Ld \bM , \bLambda _{\varphi } \Rd \frac{1}{1-\eta \, \varphi _{1} } + \Eta \, \Omega_{\varphi _{1}} \, , 
\hspace{5mm}  
\delta \varphi _{2} = \Eta \, \Omega_{\varphi _{2}} \, , 
\hspace{5mm}  
\delta \psi = \Eta \, \Omega _{\psi } \, , 
\end{align*}
where $\Lambda _{\varphi }$, $\Omega _{\varphi _{a}}$, and $\Omega _{\psi }$ denote appropriate gauge parameters. 
Hence, one can obtain the gauge invariant action (\ref{switched BV}) by replacing a part of (\ref{Berkovits BV}) by $S_{2}$, in which $\bM$-terms turn into $\Eta$-terms while keeping the gauge invariance. 
We impose the constraint equations $\widehat{\gamma } = \{ \Gamma _{g,p} , \gamma _{g,p} \} _{g,p}$, 
\begin{align}
\label{simple constraint}
\Gamma _{-1-g,-p} \equiv (\Phi _{-g,p}^{\, \xi } )^{\ast } - \eta \, \Phi _{1+g, -p} \, , 
\hspace{5mm} 
\gamma _{-1-g,-p} \equiv (\Phi _{-g,p}^{\, \eta } )^{\ast } - \xi \, \eta \, \Psi _{2+g,-1-p} \, , 
\end{align}
for each spacetime ghost number. 
Note that $(\Gamma , \gamma )=0$ by construction. 
We find $( \Gamma , \Gamma )^{-1} = \xi $ and $(\gamma , \gamma )^{-1}= \eta \, \xi $\,, and thus $( S_{1} , S_{1})_{\Gamma } = 0$ holds as (\ref{constrained master eq}). 
Since $(S_{2} , S_{2} )_{\gamma } = 0$ holds as (\ref{invariant constraint}), we find that $S_{\textsf{bv}}=S_{1} + S_{2}$ gives a solution of the constrained master equation, 
\begin{align*}
\big{(} \, S_{\textsf{bv}} \, , \, S_{\textsf{bv}} \, \big{)}_{\Gamma , \gamma } = 
\big{(} \, S_{1} \, , \, S_{1} \, \big{)}_{\Gamma } 
+ \big{(} \, S_{2} \, , \, S_{2} \, \big{)}_{\gamma } 
= 0 \, .  
\end{align*}
The relation between (\ref{Berkovits BV}) and (\ref{switched BV}) may be understood as a BV canonical transformation.\footnote{In the context of the conventional BV approach for the free theory, this type of $Q$-$\eta$ switching operation is just a result of BV canonical transformations. See \cite{Matsunaga:2017phm} for details. }

\section{Conventional BV approach revisited}

We gave several solutions of constrained master equations in the previous section, in which $S_{\textsf{bv}}|_{\widehat{\gamma }}$ of (\ref{switched BV}) will be rather plain. 
We can rewrite the constraints (\ref{simple constraint}) into the simple form 
\begin{align}
\widehat{\gamma } = \varphi ^{\ast } - \psi \, , 
\end{align}
where (\ref{shifted extra ghost}) is extended for all $g$ by using $\psi ^{\eta } \equiv \eta \, \varphi $\,. 
This simple expression of the constraints resembles us the conventional BV approach, and it suggests that one could construct a BV master action $S_{\textsf{bv}}$ based on the minimal set within the conventional BV approach. 

Besides reassembling string antifields, to split string fields $\varphi = \varphi _{1} + \varphi _{2}$ and to utilise each $\varphi _{a}$ as an argument of $S_{\textsf{bv}}$ play a crucial role in the constrained BV approach. 
As we show in this section, one can construct a conventional BV master action $S_{\textsf{bv}}$ as a function of $(\Phi ^{\xi })^{\ast }$ and $(\Phi ^{\eta })^{\ast }$, not a function of the sum $(\Phi )^{\ast } = (\Phi ^{\xi })^{\ast } + (\Phi ^{\eta })^{\ast }$\,. 
While we introduce the string antifield $(\Phi _{-g,p})^{\ast }$ for the string field $\Phi _{-g,p}$ as the usual conventional BV approach, we consider their $\xi $- or $\eta$-exact components separately. 
Note that $(\Phi ^{\ast })^{\eta } = (\Phi ^{\xi })^{\ast }$ and $(\Phi ^{\ast })^{\xi } = (\Phi ^{\eta })^{\ast }$ because of 
\begin{align*}
\la (\Phi _{-g,p}^{\,\, \xi } )^{\ast } , \, \Phi _{-g',p'}^{\,\, \xi } \ra = \delta _{g,g'} \, \delta _{p,p'} \, ,
\hspace{5mm} \la (\Phi _{-g,p}^{\,\, \eta } )^{\ast } , \, \Phi _{-g',p'}^{\,\, \eta } \ra = \delta _{g,g'} \, \delta _{p,p'} \, . 
\end{align*}

\subsection{Orthogonal decomposition} 

We consider the orthogonal decomposition of the gauge transformation $\delta \Phi = \delta \Phi ^{\xi } + \delta \Phi ^{\eta }$. 
By redefining gauge parameters as follows 
\begin{align}
\label{redef} 
\Lambda _{-1,0}^{\textrm{new}} 
\equiv - \Lambda _{-1,0}^{\textrm{old}}  \, , 
\hspace{5mm}  
\Lambda _{-1,1}^{\textrm{new}}
\equiv  \pi _{1} \, \bxi \, \Ld \, \bM , \, \bLambda _{-1,0}^{\textrm{old}} \, \Rd \frac{1}{1-\eta \, \Phi } + \Lambda _{-1,1}^{\textrm{old}} \, , 
\end{align}
we set the $\eta $-exact component of the gauge transformations $\delta \Phi ^{\eta }$ linear 
\begin{subequations}
\begin{align}
\label{pgf a}
\delta \Phi ^{\xi } & = \pi _{1} \, \bxi \, \Ld \, \bM , \, \Eta \, \bLambda _{-1,0}^{\textrm{new}} \, \Rd \frac{1}{1-\eta \, \Phi } \, , 
\\ \label{pgf}
\delta \Phi ^{\eta } & = \Eta \, \Lambda _{-1,1}^{\textrm{new}} \, . 
\end{align} 
\end{subequations}
As we will see, it enables us to simplify the other higher gauge transformations: 
Except for the gauge parameters $\{ \Lambda _{-g,0} \} _{g>0}$ carrying picture number $0$, the $\xi $-components of $\delta \Lambda _{-g,p} $ are proportional to the equations of motion and thus become trivial transformations; 
The $\eta$-exact components of all $\delta \Lambda _{-g,p}$ can be linearised by redefining the gauge parameters. 
We find 
\begin{align*}
\delta \Lambda _{-1,1}^{\textrm{new}} & = \pi _{1} \, \bxi \, \Ld \, \bM , \, \delta \bLambda _{-1,0}^{\textrm{old}} \, \Rd \frac{1}{1-\eta \, \Phi } + \delta \Lambda _{-1,1}^{\textrm{old}} 
\no 
& = \pi _{1} \, \bxi \, \LD \, \bM , \, \pi _{1} \Ld \bM , \bLambda _{-2,1}^{\textrm{old}} \Rd \frac{1}{1-\eta \Phi } \RD \frac{1}{1-\eta \, \Phi } 
+ \Eta \, \bigg[ \pi _{1} \, \bxi \, \Ld \bM , \bLambda _{-2,1}^{\textrm{old}} \Rd \frac{1}{1-\eta \Phi } + \Lambda _{-2,2}^{\textrm{old}} \bigg] \,
\no & 
= \bxi \, T (\Lambda _{-2,1}^{\textrm{new}} ) + \Eta \, \Lambda _{-2,2}^{\textrm{new}} \, , 
\end{align*}
where $T(\Lambda ) \equiv \pi _{1} \Ld \ld \bM , \bLambda \rd , (\textrm{e.o.m}) \Rd \frac{1}{1-\eta \Phi }$ denotes a trivial transformation. 
Hence, for any $0<p\leq g$, the higher gauge transformations can be rewritten as follows 
\begin{subequations} 
\begin{align}
\delta \Lambda _{-g,0}^{\textrm{new}} & = \pi _{1} \, \bxi \, \ld \bM , \Eta \, \bLambda _{-g-1,0}^{\textrm{new}} \rd \frac{1}{1 - \eta \, \Phi } + \Eta \, \Lambda _{-g-1,1}^{\textrm{new}}  \, , 
\\ \label{pgf higher} 
\delta \Lambda _{-g,p}^{\textrm{new}} & =  \Eta \, \Lambda _{-g-1,p+1}^{\textrm{new}} \, . 
\end{align}
\end{subequations} 
While the orthogonal decomposition $\delta\Phi = \delta \Phi ^{\xi } + \delta \Phi ^{\eta }$ makes $\delta \Lambda ^{\xi }$ trivial for $p>0$, redefinitions of $\Lambda $ make $\delta \Lambda ^{\eta }$ linear. 
These operations enable us to obtain a simple BV master action. 

\vspace{1mm}

Note that partial gauge fixing is an operation omitting $\Phi ^{\eta }$ and (\ref{pgf}) at the classical level. 
Then, the line of $\Lambda _{-g,p=g}$ of (\ref{pgf higher}) does not appear in its higher gauge transformations. 
It gives the gauge reducibility of partially gauge-fixed superstring field theory in the large Hilbert space, in which reassembled string fields--antifields which correspond to (\ref{redef}) will be rather natural. 

\subsection{BV master action} 

Let $F$ be a functional of the minimal set of spacetime fields--antifields, which may be a functional of string fields or string antifields. 
We perturbatively construct $S_{\textsf{bv}}$ satisfying 
\begin{align*}
\delta _{\textsf{bv}} F  = \big{(} \, F \, , \, S_{\textsf{bv}} \, \big{)}_{\textrm{min}} \, , 
\end{align*}
whose nilpotency is our guiding principle. 
The initial condition is $S_{\textsf{bv}}|^{(0)} = S[\Phi ]$. 
We write $(\Phi ^{\xi })^{\ast }$ or $(\Phi ^{\eta } )^{\ast }$ for the string antifield for the $\xi $- or $\eta$-exact component of the dynamical string field $\Phi = \Phi ^{\xi }+\Phi ^{\eta }$, respectively. 
As we will see, $S_{\textsf{bv}}$ becomes a functional of $(\Phi ^{\xi })^{\ast }$ and $(\Phi ^{\eta })^{\ast }$\,. 
We require that as well as $\Phi ^{\xi } \in \textrm{Im} [\xi ]$ and $\Phi ^{\eta } \in \textrm{Im}[\eta ]$, their BV transformations satisfy 
\begin{align*}
\delta _{\textsf{bv}} \Phi ^{\xi } = \big{(} \, \Phi ^{\xi } , \, S_{\textsf{bv}} \, \big{)}_{\textrm{min}} = \frac{\partial S_{\textsf{bv}} }{\partial (\Phi ^{\xi })^{\ast } } \in \textrm{Im} [ \xi ] 
\, , \hspace{5mm} 
\delta _{\textsf{bv}} \Phi ^{\eta } = \big{(} \, \Phi ^{\eta } , \, S_{\textsf{bv}} \, \big{)}_{\textrm{min}} = \frac{\partial S_{\textsf{bv}} }{\partial (\Phi ^{\eta })^{\ast } } \in \textrm{Im} [ \eta ]
\, . 
\end{align*}
Further, we require that the $\eta $-exact component of the BV transformations are linear 
\begin{align}
\label{linear} 
\delta _{\textsf{bv}} \Phi _{-g,p}^{\, \eta } = \big{(} \, \Phi _{-g,p}^{\, \eta } \, , \, S_{\textsf{bv}} \, \big{)}_{\textrm{min}} = \Eta \, \Phi _{-g-1,p+1} \, .  
\end{align} 
In other words, we consider redefinitions of gauge parameter fields given in section 6.1 and focus on the gauge algebra of the orthogonally decomposed gauge transformations. 
In general, $\delta \Phi _{-g,p}^{\eta }$ could be a nonlinear function of fields--antifields.
This requirement (\ref{linear}) is too restrictive and should be removed to find a more general form of the BV master action in the large Hilbert space.  
However, as we will see, this requirement prohibits any interacting terms of $\Phi _{-g,p \not= 0}$ or $(\Phi _{-g,p\not= 0})^{\ast }$, and it enables us to construct a simple BV master action. 
We find that string-antifield derivatives of $S^{(1)}$ are given by 
\begin{subequations}
\begin{align} 
\label{afd a}
\delta _{\textsf{bv}} \Phi ^{\xi } |^{(0)} = \big{(} \,  \Phi ^{\xi } \, , \, S_{\textsf{bv}} \, \big{)} |^{(0)} = 
\frac{\partial S^{(1)} }{\partial (\Phi ^{\xi } )^{\ast } } & = \pi _{1} \, \bxi \, \Ld \bM , \Eta \, \bPhi _{-1,0} \Rd \frac{1}{1-\eta \, \Phi } \, , 
\\
\delta _{\textsf{bv}} \Phi ^{\eta } |^{(0)} = \big{(} \, \Phi ^{\eta } \, , \, S_{\textsf{bv}} \, \big{)} |^{(0)} =
\frac{\partial S^{(1)} }{\partial (\Phi ^{\eta } )^{\ast } } & =  \Eta \, \Phi _{-1,1} \, , 
\end{align}
\end{subequations} 
which are determined from the gauge transformations (\ref{pgf a}-b)  and their gauge algebra. 
Note that $(\Phi ^{\xi })^{\ast }$ is $\eta $-exact $(\Phi ^{\xi} )^{\ast } = \cP ^{\eta } (\Phi ^{\xi} )^{\ast }$ and $(\Phi ^{\eta })^{\ast }$ is $\xi$-exact $(\Phi ^{\eta } )^{\ast } = \cP ^{\xi } (\Phi ^{\eta } )^{\ast }$\,. 
These string-antifield derivatives (\ref{afd a}-b) determine the antifield number $1$ part of $S_{\textsf{bv}}$ as follows 
\begin{align}
\label{afn 1 part}
S^{(1)} = \La ( \Phi ^{\xi } )^{\ast } , \, \bxi \, \Ld \bM , \Eta \, \bPhi _{-1,0} \Rd \frac{1}{1-\eta \, \Phi } \Ra 
+ \la ( \Phi ^{\eta } )^{\ast } , \, \Eta \, \Phi _{-1,1} \ra \, . 
\end{align}
Note that this $S^{(1)}$ is not a functional of $\Phi ^{\ast } = (\Phi ^{\xi })^{\ast } + (\Phi ^{\eta })^{\ast }$ but a functional of $(\Phi ^{\xi })^{\ast }$ and $(\Phi ^{\eta })^{\ast }$. 
Clearly, string-field derivatives of (\ref{afn 1 part}) become $\eta $-exact states as follows 
\begin{align*}
& \hspace{15mm} 
\frac{\partial S^{(1)} }{\partial \Phi ^{\xi } } = \pi _{1} \, \Ld \ld \bM , \Eta \, \bPhi _{-1,0} \rd , (\bPhi ^{\xi })^{\ast } \Rd \frac{1}{1-\eta \, \Phi } \, ,  
\\ & 
\frac{\partial S^{(1)} }{\partial \Phi _{-1,0}^{\,\, \xi } } = \pi _{1} \,  \Ld \bM , (\bPhi ^{\xi })^{\ast } \Rd \frac{1}{1-\eta \, \Phi } \, , 
\hspace{12mm}  
\frac{\partial S^{(1)} }{\partial \Phi _{-1,1}^{\,\, \xi } } = \Eta \, ( \Phi ^{\eta } )^{\ast } \, . 
\end{align*}
In other words, as the original action, a half of the string-field derivatives vanish: 
\begin{align*}
\frac{\partial S^{(1)} }{\partial \Phi ^{\eta } } = \frac{\partial S^{(1)} }{\partial \Phi _{-1,0}^{\,\, \eta } } = \frac{\partial S^{(1)} }{\partial \Phi _{-1,1}^{\,\, \eta } }  = 0  
\hspace{8mm} \textrm{as} \hspace{5mm} 
\frac{\partial S}{\partial \Phi ^{\eta } }= 0 \, . 
\end{align*}
By construction of (\ref{linear}), a half of the string-antifield derivatives of $S^{(2)}$ are given by  
\begin{align*}
\delta _{\textsf{bv}} \Phi ^{\eta } |^{(1)} & = 
\frac{\partial S^{(2)} }{\partial (\Phi ^{\eta })^{\ast } } = 0 \, , 
\hspace{8mm}  
\delta _{\textsf{bv}} \Phi _{-1,p}^{\,\, \eta } |^{(0)} = 
\frac{\partial S^{(2)} }{\partial (\Phi _{-1,p}^{\,\, \eta })^{\ast } } =\Eta \, \Phi _{-2,1+p} \, 
\hspace{3mm}  (p = 0,1 ) \, . 
\end{align*}
To solving the master equation, the other string-antifield derivatives of $S^{(2)}$ have to take  
\begin{align*}
\delta _{\textsf{bv}} \Phi ^{\xi } |^{(1)} & =  
\frac{\partial S^{(2)} }{\partial (\Phi ^{\xi })^{\ast } } = \pi _{1} \, \bxi \LD 
\Ld \ld \bM , \Eta \, \bPhi _{-1,0} \rd , \Eta \, \bPhi _{-1,0} \Rd 
+ \ld \bM , \Eta \, \bPhi _{-2,0} \rd 
, (\bPhi ^{\xi })^{\ast } \RD \frac{1}{1-\eta \, \Phi } \, , 
\\
\delta _{\textsf{bv}} \Phi _{-1,0}^{\,\, \xi } |^{(0)} & = 
\frac{\partial S^{(2)} }{\partial (\Phi _{-1,0}^{\,\, \xi })^{\ast } } 
= \pi _{1} \, \bxi \, \bigg[ \Ld \ld \bM , \Eta \, \bPhi _{-1,0} \rd ,  \Eta \, \bPhi _{-1,0}  \Rd + \ld \bM , \Eta \, \bPhi _{-2,0} \rd \bigg] \frac{1}{1-\eta \, \Phi }  \, , 
\\
\delta _{\textsf{bv}} \Phi _{-1,1}^{\,\, \xi } |^{(0)} & = 
\frac{\partial S^{(2)} }{\partial (\Phi _{-1,1}^{\,\, \xi })^{\ast } } = 0 \, . 
\end{align*}
Note that the requirement (\ref{linear}) prohibits not only nonlinear $\eta$-transformations but also the interacting terms of $\Phi _{-1,1}$\,. 
These derivatives determine the antifield number $2$ part of the master action $S^{(2)}$ satisfying $\big{(} S^{(0)} + S^{(1)} + S^{(2)} + \cdots , \, S^{(0)} + S^{(1)} + S^{(2)} + \cdots \big{)}_{\textrm{min}} = 0$. 
Likewise, one can construct $S^{(3)}$, $S^{(4)}$, and higher $S^{(n)}$ on the basis of the antifield number expansion. 
These are functionals of $\Phi _{-g,p}$, $(\Phi _{-g,p}^{\, \xi })^{\ast }$, and $(\Phi _{-g,p}^{\, \eta })^{\ast }$ as expected.

\vspace{1mm}

Let $\varphi _{p}$ be the sum of all string fields carrying world-sheet picture number $p$, which can be decomposed as $\varphi _{p} = \varphi _{p}^{\xi } + \varphi _{p}^{\eta }$\,. 
We write $(\varphi _{p}^{\xi } )^{\ast }$ or $(\varphi _{p}^{\eta })^{\ast }$ for the string antifield for the $\xi $- or $\eta $-exact component of $\varphi _{p}$ respectively as follows, 
\begin{align*} 
\varphi _{p} \equiv \sum_{g=p}^{\infty } \Phi _{-g,p} \, ,
\hspace{5mm} 
(\varphi _{p}^{\xi } )^{\ast } = \sum_{g=p}^{\infty } ( \Phi _{-g,p}^{\, \xi } )^{\ast }  \, , 
\hspace{5mm} 
(\varphi _{p}^{\eta } )^{\ast } = \sum_{g=p}^{\infty } ( \Phi _{-g,p}^{\, \eta } )^{\ast }  \, . 
\end{align*}
The dynamical string field $\Phi$ is included in $\varphi _{0}$ and the sum $\varphi $ of all string fields is given by $\varphi = \varphi _{0} + \sum_{p>0} \varphi _{p}$\,. 
We find that the BV master action $S_{\textsf{bv}} = S_{\textsf{bv}} [ \, \varphi , ( \varphi ^{\xi })^{\ast } ( \varphi ^{\eta } )^{\ast } ]$ is 
\begin{align}
\label{revisited}
S_{\textsf{bv}} 
= \int_{0}^{1} dt \, \LA \varphi _{0} + \xi \, ( \varphi _{0}^{\xi } )^{\ast } , \, \bM \frac{1}{1- t \, \eta \, ( \varphi _{0} + \xi \, ( \varphi _{0}^{\xi } )^{\ast } ) } \RA + \sum_{p > 0} \La ( \varphi _{p-1}^{\eta } )^{\ast } , \, \Eta \, \varphi _{p} \Ra \, . 
\end{align} 
While the first term is a functional of $\varphi _{0}$ and $(\varphi _{0}^{\xi })^{\ast}$, the second term is a functional of $\varphi _{p>0}$, $(\varphi _{0}^{\eta })^{\ast }$, and $(\varphi _{p}^{\eta })^{\ast }$. 
The variation of $S_{\textsf{bv}}$ takes the following form  
\begin{align*}
\delta S_{\textsf{bv}} & = \LA \, \delta \varphi _{0} , \, \bM \frac{1}{1- \eta \, ( \varphi _{0} + \xi \, ( \varphi _{0}^{\xi } )^{\ast } ) } \RA 
+ \sum_{p>0} \La \delta \varphi _{p} , \, \Eta \, ( \varphi _{p-1}^{\, \eta } )^{\ast } \Ra
\no & \hspace{10mm}
+ \LA \, \delta ( \varphi _{0}^{\xi } )^{\ast } , \, \bxi \, \bM \frac{1}{1-\eta \, ( \varphi _{0} + \xi \, ( \varphi _{0}^{\xi } )^{\ast } ) } \RA 
+ \sum _{p>0} \La \delta ( \varphi _{p-1}^{\, \eta } )^{\ast } , \, \Eta \, \varphi _{p} \Ra \, . 
\end{align*}
Note that $\Phi _{-g,p}$ for $p>0$ has no interacting term, and thus the third term has no contraction with the second or fourth term in the master equation. 
Clearly, our master action (\ref{revisited}) satisfies 
\begin{align*}
\frac{1}{2} 
\big{(} \, S_{\textsf{bv}} , \, S_{\textsf{bv}} \, \big{)}_{\textrm{min}} 
= \frac{\overset{\leftarrow }{\partial } S_{\textsf{bv}} }{\partial \varphi ^{\xi }} \cdot \frac{\overset{\rightarrow }{\partial } S_{\textsf{bv}} }{\partial (\varphi ^{\xi } )^{\ast }} 
+ \frac{\overset{\leftarrow }{\partial } S_{\textsf{bv}} }{\partial \varphi ^{\eta }} \cdot \frac{\overset{\rightarrow }{\partial } S_{\textsf{bv}} }{\partial ( \varphi ^{\eta } )^{\ast }} = 0 \, .
\end{align*}
While the BV transformations of string fields take the following forms, 
\begin{subequations}
\begin{align}
\delta \varphi _{0} & = \big{(} \, \varphi _{0}^{\xi } + \varphi _{0}^{\eta }  \, , \, S_{\textsf{bv}} \, \big{)}_{\textrm{min}} 
= \pi _{1} \, \bxi \, \bM \frac{1}{1 - \eta \, ( \varphi _{0} + \xi \, (\varphi _{0} )^{\ast }) } + \Eta \, \varphi _{1}  \, , 
\\
\delta \varphi _{p} & = \big{(} \, \varphi _{p}^{\xi } + \varphi _{p}^{\eta } \, , \, S_{\textsf{bv}} \, \big{)}_{\textrm{min}} 
= \Eta \, \varphi _{p+1} \, , 
\end{align}
\end{subequations} 
the BV tramsformations of string antifields are given by 
\begin{subequations}
\begin{align} 
\delta (\varphi _{0} )^{\ast } & = \big{(} \, (\varphi _{0}^{\xi })^{\ast } + (\varphi _{0}^{\eta })^{\ast } \, , \, S_{\textsf{bv}} \, \big{)}_{\textrm{min}} = 
\pi _{1} \, \bM \frac{1}{1 - \eta \, ( \varphi _{0} + \xi \, (\varphi _{0} )^{\ast }) } \, , 
\\
\delta (\varphi _{p} )^{\ast } & = \big{(} \, (\varphi _{p}^{\xi })^{\ast } + (\varphi _{p}^{\eta })^{\ast }  \, , \, S_{\textsf{bv}} \, \big{)}_{\textrm{min}} = \Eta \, ( \varphi _{p-1}^{\, \eta } )^{\ast } \, . 
\end{align}
\end{subequations} 
The role of the second term of (\ref{revisited}) in perturbation theory depends on the gauge-fixing condition: 
For example, it is integrated out and trivially decouples in the Siegel gauge; however, it will provide nontrivial contributions to loop amplitudes in the $d_{0}$-gauge. 
See \cite{Kroyter:2012ni, Torii:2012nj}.

\section{Concluding remarks} 

In this paper, we developed the Batalin-Vilkovisky formalism of superstring field theory in the large Hilbert space with the goal of understanding of how to construct large master actions for interacting theory. 
We first showed that the constrained BV approach \cite{Batalin:1992mk} is well applicable, in which Berkovits' simple prescription \cite{Berkovits:2012np} is rather suitable for the large but partially gauge-fixed theory. 
By modifying its constraints, extra string fields, or starting unconstrained action, we constructed several constrained BV master actions in the large Hilbert space. 
We next showed that the conventional BV approach is also applicable iff we give up constructing master actions as naive functionals of string fields--antifields. 
We constructed a BV master action as a functional of fine parts $\{ \varphi _{a} \}_{a=1}^{n}$ of string fields--antifields $\varphi = \varphi _{1} + \cdots + \varphi _{n}$, not a function of string fields--antifields themselves. 
It is worth mentioning that our analysis is quickly applicable to the \textit{large} theory which is obtained by embedding \cite{Erler:2016ybs} or \cite{Erler:2014eba}, and thus BV master actions for the \textit{large} $A_{\infty }$ theory including the Ramond sector or the \textit{large} $L_{\infty }$ theory are constructed in the completely same manner. 
Also, since BV master actions in the large Hilbert space are obtained, one can discuss the validity of partial gauge-fixing now. 
We conclude with some remarks. 

\vspace{-2.5mm}

\subsubsection*{BV formalism in the large Hilbert space} 

\vspace{-1.5mm} 

First, it is worth visiting and connecting different pairs of $(S_{\textsf{bv}} , \widehat{\Gamma } , \varphi _{\textsf{ex}} )$ to obtain a better understanding of the BV formalism in the large Hilbert space. 
While we gave several constrained BV master actions in section 5, there would exist some canonical transformations connecting these. 
Next, it is desirable to find a more general form of the master action in the large Hilbert space on the basis of the conventional BV approach. 
Our master action (\ref{revisited}) has a simple form but is constructed based on the requirement (\ref{linear}), which will be too restrictive. 

\vspace{-2.5mm} 

\subsubsection*{WZW-like formulation} 

\vspace{-1.5mm} 

Our results give a simple example of constructing BV master actions for WZW-like superstring field theory (\ref{S}), which is based on the parametrisation (\ref{another sol}). 
It is important to clarify whether one can construct a BV master action for other parametrisation of the WZW-like functional (\ref{pure}), such as $A_{\eta }[\Phi ] = ( \eta e ^{\Phi } ) e^{- \Phi }$. 
One may be able to apply constrained BV approach for the Berkovits theory in a similar manner, in which partial gauge fixing may take a nonlinear form such that $(\delta e^{\Phi }) e^{-\Phi }= Q \Lambda + \eta \, \Omega - \ld A_{\eta } , \Omega \rd $ is orthogonally decomposed as (\ref{pgf a}-b). 
A procedure which does not depend on these parametrisations would be necessitated to apply the BV formalism to the general WZW-like formulation \cite{Matsunaga:2016zsu, Erler:2017onq}. 

\vspace{-2.5mm} 

\subsubsection*{Gauge tensors' formulae} 

\vspace{-1.5mm} 

Since our master actions are not naive functionals of string fields--antifields, it is interesting to clarify the difference of 
the gauge tensors based on spacetime and string fields. 
It reveals why a ready-made BV procedure does not work in the large Hilbert space from the point of view of the gauge algebra. 
Then, how ``partial gauge fixing'' fixes the gauge would be clarified. 

\section*{Acknowledgements} 
The authors would like to thank Mitsuhiro Kato and the organizers of "Strings, Fields, and Particles 2017 at Komaba". 
H.M. also thanks Ted Erler, Hiroshi Kunitomo, Yuji Okawa, Martin Schnabl, and Shingo Torii. 
M.N. is grateful to Yuji Tachikawa. 
This research has been supported by the Grant Agency of the Czech Republic, under the grant P201/12/G028.

\appendix
\section{Notations and basic identities} 
\vspace{-1mm} 

In this appendix, we explain our notation and some basic identities briefly. 
See also \cite{A infinity and bv, Erler:2015uoa, Matsunaga:2017phm}. 
In general, a string field $\Psi \equiv \psi \, \cZ $ consists of a set of spacetime fields $\psi $ having spacetime ghost number and a set of world-sheet bases $\cZ $ which carry world-sheet ghost and picture numbers. 
Its state space $\cH $ is a vector space, which is equipped with the BPZ inner product or graded symplectic structure.
String products $\bM $ define a linear map acting on its tensor algebra $\cT (\cH )$. 

\subsection*{On the BPZ inner product}
\vspace{-1mm}

Let $s [ \varphi ]$, $\mathrm{gh} [ \varphi ]$, and $\mathrm{pc} [\varphi ]$ be the spacetime ghost, world-sheet ghost, and picture numbers of $\varphi $, respectively. 
The Grassmann parity of $\varphi $ is the sum of the spacetime and world-sheet ghost numbers, which we write $\mathrm{G} [ \varphi ] \equiv s [ \varphi ] + \mathrm{gh} [ \varphi ]$\,. 
The upper index of $(-)^{\varphi }$ denotes the $\varphi$'s Grassmann parity $(-)^{\varphi } \equiv (-)^{\mathrm{G}[\varphi ]}$ for brevity, if $(-)^{\varphi }$ is put on the BPZ inner product. 
We write ${\textrm{G}}[\omega ]$ for the grading of the BPZ inner product; it is $0$ in the large Hilbert space; it is $1$ in the small Hilbert space. 
The BPZ inner product $\langle \varphi _{1} , \varphi _{2} \rangle _{\textsf{bpz}}$ is graded symmetric and bilinear 
\begin{subequations} 
\begin{align}
\label{graded sym}
\la \, \Psi \, , \, \Phi \, \ra _{\textsf{bpz}} & = (-)^{ \Psi \, \Phi } \la \, \Phi \, , \, \Psi \, \ra _{\textsf{bpz}} \, , 
\\  
\la \, \psi \, \cZ \, , \, \Phi \, \ra _{\textsf{bpz}} & = (-)^{ {\textrm{G}} [ \omega ] {\textrm{G}} [ \psi ] } \psi  \, \la \, \cZ \, , \, \Phi \, \ra _{\textsf{bpz}} \, , 
\\ 
\la \, \Phi \, , \, \psi \, \cZ \, \ra _{\textsf{bpz}} & = (-)^{{\textrm{G}} [\psi  ] {\textrm{G}} [\cZ ] } \la \, \Phi \, , \, \cZ \, \ra _{\textsf{bpz}} \, \psi  \, .
\end{align}
\end{subequations}
We focus on the large Hilbert space. 
There exists a complete system of BPZ bases satisfying 
\begin{align}
\label{BPZ basis}
\la \, \cZ _{g,p}^{a} \, , \, \cZ _{h, q}^{b} \, \ra _{\textsf{bpz}} 
= (-)^{\textrm{gh}[\cZ _{h,q} ] \textrm{pc} [\cZ _{h,q}] } 
\delta ^{a,b} \, \delta _{g+h, 2} \, \delta _{p+q , -1} \, ,  
\end{align} 
where the $a$-label distinguishes different bases carrying the same world-sheet ghost and picture numbers.  
Note that the $g$-label can run over all integer numbers\footnote{\textbf{Torii's subset:} In the earlier works \cite{Torii:2011zz, Kroyter:2012ni, Torii:2012nj}, just a subset $\cB _{g,p} = \{ \cZ _{g,p} , \, \cZ _{g,p}^{\ast } \}_{g<1}$ is used to give the free master action. 
Because of $\cB _{g,p}^{a} \equiv \cZ _{g,p}^{a}$ for $g<1$ and $\cB _{g,p}^{a} \equiv \cZ _{g,p}^{a \, \ast }$ for $g>1$, it satisfies 
\begin{align*}
\la \cB _{g,p}^{a} , \, \cB _{h,q}^{b} \ra _{\textsf{bpz}} =  
\big[ \theta ( 1>g ) + (-)^{g} \theta (g > 1) \big] 
\delta ^{a,b} \, \delta _{g+h, 2} \, \delta _{p+q , -1} \, , 
\end{align*} 
where $\theta (g>1)$ is a step function of $g$. 
This subset satisfies (\ref{1>g}) for $1>g$ and (\ref{g>1}) for $g>1$\,. } 
because of $p+q=-1$ and these bases indeed satisfy (\ref{graded sym}). 
By absorbing the sign of (\ref{BPZ basis}), one can define  
\begin{align} 
\label{dual BPZ basis}
\cZ _{2-g,-1-p}^{a \, \ast } \equiv  (-)^{ \textrm{gh} [\cZ _{g,p} ] \textrm{pc} [\cZ _{g,p}] } \cZ _{g,p}^{a} \, , 
\end{align}
which satisfy simpler relations $\langle \cZ _{g,p}^{a} , \cZ _{h, q}^{b \, \ast } \rangle _{\textsf{bpz}} = (-)^{\textrm{gh} [\cZ _{g,p}^{\ast }] } \langle \cZ _{g,p}^{a\, \ast } , \cZ _{h, q}^{b} \rangle _{\textsf{bpz}} = \delta ^{a,b} \, \delta _{g,h} \, \delta _{p,q} $\,. 
These dual bases are useful to give string antifields. 
We often omit the $a$-label for simplicity.

\vspace{1mm} 

The orthogonal relation (\ref{dual BPZ basis}) provides simple decompositions of the unit. 
We introduce the symbol $\delta (r)$ satisfying $\sum _{r} \delta (r) = 1$. 
From (\ref{BPZ basis}), we obtain 
$\langle \cZ _{g,p}^{a} , \cZ _{f, r}^{c} \rangle _{\textsf{bpz}} 
\langle \cZ _{2-f,-1-r}^{c} , \cZ _{h, q}^{b} \rangle _{\textsf{bpz}} 
= (-)^{fr} \delta ^{a,c} \delta _{g+f,2} \delta _{p+r,-1} \cdot (-)^{h,q} \delta ^{c,b} \delta _{f,h,} \delta _{r,q}$\,. 
By summing over $(f,r)$ and using the definition of the dual bases (\ref{dual BPZ basis}), we find 
\begin{subequations} 
\begin{align}
\label{1>g}
\la \cZ _{g,p}^{a} , \, \cZ _{f, r}^{c \, \ast } \ra _{\textsf{bpz}} 
\la \cZ _{f,r}^{c} , \, \cZ _{h, q}^{b} \ra _{\textsf{bpz}} 
& = \la \cZ _{g,p}^{a} , \, \cZ _{h,q}^{b} \ra _{\textsf{bpz}}  
\delta (c) \delta (f) \delta (r) \, , 
\\ \label{g>1}
(-)^{ \textrm{gh} [ \cZ _{f,r}^{\ast } ] }
\la \cZ _{g,p}^{a} , \, \cZ _{f, r}^{c} \ra _{\textsf{bpz}} 
\la \cZ _{f,r}^{c \, \ast } , \, \cZ _{h, q}^{b} \ra _{\textsf{bpz}} 
& = \la \cZ _{g,p}^{a} , \, \cZ _{h,q}^{b} \ra _{\textsf{bpz}}  
\delta (c) \delta (f) \delta (r) \, . 
\end{align}
\end{subequations}  
Utilizing ${}_{\textsf{b}} \langle A | B \rangle _{\textsf{b}} = \langle A , B \rangle _{\textsf{bpz}}$\,, these relations can be expressed as follows 
\begin{align*}
\sum _{g,p,a } | \cZ _{g,p}^{a \, \ast } \rangle _{\textsf{b}} \, {}_{\textsf{b}} \langle \cZ _{g,p}^{a} | = 1 \, , 
\hspace{5mm}
\sum _{g,p,a } (-)^{\textrm{gh} [ \cZ _{g,p}^{\ast } ] } | \cZ _{g,p}^{a} \rangle _{\textsf{b}} \, {}_{\textsf{b}} \langle \cZ _{g,p}^{a \, \ast } | = 1 \, . 
\end{align*}

\subsection*{On the graded symplectic form}
\vspace{-1mm} 

We write $\mathrm{deg}[\varphi ]$ for the world-sheet degree of $\varphi $, which is defined by $\mathrm{deg} [\varphi ] \equiv \mathrm{gh}[ \varphi ] - 1$\,. 
Then, the (total) degree $\epsilon [\varphi ]$ of $\varphi $ is defined by $\epsilon [\varphi ] \equiv \mathrm{deg} [\varphi ] + s [\varphi ]$\,. 
Note that spacetime fields $\psi $ has no world-sheet degree $\mathrm{deg} [\psi ] = 0$ and thus its (total) degree is equal to its spacetime ghost number $\epsilon [\psi ] = s[\psi ]$\,. 
Since the BPZ inner product is graded symmetric, its suspension gives the graded symplectic form 
\begin{align} 
\label{susp}
\la \, \Psi \, , \, \Phi \, \ra \equiv - (-)^{\mathrm{gh} [ \Psi ] + s [ \Phi ] } \la \, \Psi \, , \, \Phi \, \ra _{\mathsf{bpz}} \, .
\end{align}
Note that $(-)^{\varphi }$ denotes the $\varphi$'s Grassmann parity $(-)^{\varphi } \equiv (-)^{\textrm{G} [\varphi ]}$ even if $(-)^{\varphi }$ is put on the graded symplectic form. 
By construction, we can find its defining properties 
\begin{subequations}
\begin{align}
\label{graded simp}
\la \, \Psi \, , \, \Phi \, \ra & = - (-)^{\epsilon [ \Psi ] \epsilon [ \Phi ] } \la \, \Phi \, , \, \Psi \, \ra \, ,
\\
\la \, \psi \, \cZ  , \, \Phi \, \ra & = (-)^{\epsilon [\omega ] \epsilon [ \psi ] } \psi \, \la \, \cZ \, , \, \Psi \, \ra \, , 
\\
\la \, \Phi \, , \, \psi \, \cZ \, \ra & = (-)^{\epsilon [ \psi ] \epsilon [ \cZ ] } \la \, \Phi \, , \, \cZ \, \ra \, \psi \, , 
\end{align} 
\end{subequations} 
where $\epsilon [\omega ]$ denotes the total degree of the graded symplectic form (\ref{susp}); it is $1$ in the large Hilbert space; it is $0$ in the small Hilbert space. 
Since spacetime fields has no world-sheet degree, we find $\psi \, \cZ  = (-)^{\epsilon [\psi ] ( \epsilon [ \cZ ] + 1 )} \cZ \, \psi $ and $\psi _{1} \, \psi _{2} = (-)^{\epsilon [ \psi _{1}] \epsilon [\psi _{2}] } \psi _{2} \, \psi _{1}$\,. 
We focus on the large Hilbert space. 
There exists a complete system of symplectic bases satisfying 
\begin{align*}
\la \cZ _{g,p}^{a} , \, \cZ _{h,q}^{b} \ra 
= -(-)^{\textrm{gh}[\cZ _{g,p}] \textrm{pc} [ \cZ _{g,p} ] } \delta ^{a,b} \delta _{g+h,2} \delta _{p+q,-1} \, . 
\end{align*}
Note that the $g$-label runs over all integer numbers because of (\ref{graded simp}). 
One can define their dual bases by minus (\ref{dual BPZ basis}) and quickly find $\langle \cZ _{g,p}^{a \, \ast } , \cZ _{h,q}^{b} \rangle = 
- (-)^{\mathrm{deg} [ \cZ _{g,p}^{\ast } ] }
\langle \cZ _{g,p}^{a} , \cZ _{h,q}^{b \, \ast } \rangle = - \delta ^{a,b} \delta _{g,h} \delta _{p,q}$\,. 

\vspace{1mm}

The completeness condition of the symplectic bases is given by 
\begin{align*}
- (-)^{f(r+1)} \la \cZ _{g,p}^{a} , \, \cZ _{f, r}^{c} \ra  
\la \cZ _{2-f,-1-r}^{c} , \, \cZ _{h, q}^{b} \ra  
= \la \cZ _{g,p}^{a} , \, \cZ _{h,q}^{b} \ra  \, \delta (c) \delta (f) \delta (r) \, .
\end{align*} 
By using the definition of the dual basis (\ref{dual BPZ basis}), we get symplectic versions of (\ref{1>g}) and (\ref{g>1}). 
In the sense of $\langle A | B \rangle  = \langle A , B \rangle $\,, this relation can be expressed as follows 
\begin{align*}
\sum _{g,p,a } | \cZ _{g,p}^{a} \rangle \, \langle \cZ _{g,p}^{a\, \ast } | = -1 \, , 
\hspace{5mm}
\sum _{g,p,a } (-)^{{\textrm{deg} } [\cZ _{g,p}^{\ast }] } | \cZ _{g,p}^{a \, \ast } \rangle \,  \langle \cZ _{g,p}^{a} | = 1 \, . 
\end{align*}

\subsection*{Half bases and projectors} 
\vspace{-1mm} 

We write $\cZ ^{a \, \eta }_{g,p}$ for a complete symplectic basis in the small Hilbert space satisfying  
\begin{align*}
\lla \, \cZ ^{a \, \eta }_{g,p} \, , \, \cZ ^{b \, \eta }_{h,q} \, \rra = \delta ^{a,b} \, \delta _{g+h,3} \, \delta _{p+q,-2} \, , 
\end{align*}
which is equivalent to the \textit{half} basis defined in (\ref{half}). 
By comparing the large and small bases via $\slla \cZ ^{\eta }_{g,p} , \cZ ^{\eta }_{h,q} \srra \equiv - \langle \xi \cZ ^{\eta }_{g,p} , \cZ ^{\eta }_{h,q} \rangle $, we find the following convenient relations 
\begin{align}
\label{half basis}
\big{|} \cZ _{g,p}^{\,\, \eta } \ra \equiv \eta \, \xi \, \big{|} \cZ _{g,p} \ra 
= (-)^{gp} \, \eta \, \big{|} \cZ _{g-1,p+1}^{\, \xi } \ra \, , 
\hspace{5mm} 
\big{|} \cZ _{g,p}^{\,\, \xi } \ra \equiv \xi \, \eta \, \big{|} \cZ _{g,p} \ra 
= (-)^{gp} \, \xi \, \big{|} \cZ _{g+1,p-1}^{\,\, \eta } \ra \, , 
\end{align}
where $g$ and $p$ of $(-)^{gp}$ denote world-sheet ghost and picture numbers. 
Note that $\cP _{-g,p}^{\,\, \eta } \equiv - | \cZ _{-g,p}^{\,\, \eta } \rangle ^{(1)} \, \langle \cZ _{-g,p}^{\, \ast \, \xi } |^{(2)}$ works as the projector onto the $\eta $- or $\xi $-exact components carrying world-sheet ghost number $-g$ and picture number $p$\,:
\begin{align*}
\cP _{-g,p}^{\,\, \eta } \, \varphi & = 
- \big{|} \cZ _{-g,p}^{\, \, \eta } \ra \, 
\la \cZ _{-g,p}^{\, \ast \, \xi } , \, \varphi \ra 
= - \big{|} \cZ _{-g,p}^{\, \, \eta } \ra \, 
\la \cZ _{-g,p}^{\, \ast \, \xi } , \, \Phi _{-g,p}^{\,\, \eta } \ra
= \phi _{-g,p}^{\,\, \eta } \, \big{|} \cZ _{-g,p}^{\, \, \eta } \ra 
= \Phi _{-g,p}^{\,\, \eta } \,  
\end{align*}
for $\varphi $ given by (\ref{all string fields}). 
Likewise, $\cP _{-g,p}^{\,\, \xi } \equiv (-)^{g+1} | \cZ _{-g,p}^{\,\, \xi } \rangle ^{(1)} \, \langle \cZ _{-g,p}^{\, \ast \, \eta } |^{(2)}$ works as the $\xi $-exact projector. 
Because of $\xi \, \cP^{\eta } = \xi \, (\cP ^{\eta } + \cP ^{\xi } ) = \xi$ and $\cP ^{\xi } \, \xi = ( \cP ^{\eta } + \cP ^{\xi } ) \, \xi  = \xi $, we find 
\begin{align*}
\sum_{g,p} (-)^{gp + g + p} \big{|} \cZ _{-g-1,p+1}^{\,\, \xi } \ra \la \cZ _{-g,p}^{\, \ast \, \xi } \big{|} 
& = - \xi \, \sum_{g,p} \big{|} \cZ _{-g,p}^{\,\, \eta } \ra \la \cZ _{-g,p}^{\, \ast \, \xi } \big{|} 
= \xi \, \Big[ \sum_{g,p} \cP _{-g,p}^{\,\, \eta } \Big] 
= \xi \, \cP ^{\eta } 
= \xi \, , 
\\
\sum_{g,p} (-)^{gp+g+p} \big{|} \cZ _{-g,p}^{\,\, \xi } \ra \la \cZ _{-g-1,p+1}^{\, \ast \, \xi } \big{|} 
& = \sum_{g,p} (-)^{g+1} \big{|} \cZ _{-g,p}^{\,\, \xi } \ra \la \cZ _{-g,p}^{\, \ast \, \eta } \big{|} \, \xi  
= \Big[ \sum_{g,p} \cP _{-g,p}^{\,\, \xi } \Big] \, \xi 
= \cP ^{\xi } \, \xi 
= \xi \, . 
\end{align*}
Since $\eta \, \cP ^{\xi } = \eta \, ( \cP ^{\eta } + \cP ^{\xi } ) = \eta $ and $\cP ^{\eta } \, \eta = ( \cP ^{\eta } + \cP ^{\xi } ) \, \eta = \eta $, we obtain  
\begin{align*}
\sum_{g,p} (-)^{(g+1)p}
\big{|} \cZ _{1-g,p-1}^{\,\, \eta } \ra \la \cZ _{-g,p}^{\, \ast \, \eta } \big{|} 
= \eta \, \sum_{g,p} (-)^{g+1} \big{|} \cZ _{-g,p}^{\,\, \xi } \ra \la \cZ _{-g,p}^{\, \ast \, \eta } \big{|} 
= \eta \, \Big[ \sum_{g,p} \cP _{-g,p}^{\,\, \xi } \Big] 
= \eta \, \cP ^{\xi } 
= \eta \, , 
\\
\sum_{g,p} (-)^{(g+1)p} 
\big{|} \cZ _{1-g,p-1}^{\,\, \eta } \ra \la \cZ _{-g+1,p-1}^{\, \ast \, \eta } \big{|} 
= - \sum_{g,p} \big{|} \cZ _{-g,p}^{\,\, \eta } \ra \la \cZ _{-g,p}^{\, \ast \, \xi } \big{|} \, \eta  
= \Big[ \sum_{g,p} \cP _{-g,p}^{\,\, \eta } \Big] \, \eta 
= \cP ^{\eta } \, \eta 
= \eta \, . 
\end{align*}

\subsection*{Coalgebraic notations}
\vspace{-1mm} 

A string product $\bM = \mathbf{Q} + \bM _{2} + \bM _{3} + \cdots $ consists of the BRST operator $Q$, a nonassociative three vertex $\bM_{2}$, and higher vertices $\bM _{n>2}$: 
These define multilinear maps $\{ Q , M_{n} \}_{n \geq 2}$ acting on the state space $\cH$. 
In particular, they give a cyclic $A_{\infty }$ algebra for open string field theory. 
In general, an $n$-linear map $M_{n} : \cH ^{\otimes n} \rightarrow \cH $ define a linear map $\bM _{n} : \cH ^{\otimes m} \rightarrow \cH ^{\otimes m-n+1}$ by the following Leibniz rule on tensors 
\begin{align*}
\bM _{n} : \Psi _{1} \otimes \cdot \! \cdot \! \cdot \otimes \Psi _{m}
\longmapsto  
\sum_{k=1}^{m-n} (-)^{\sigma _{k}} \Psi _{1} \otimes \cdot \! \cdot \! \cdot \otimes \Psi _{k-1} 
\otimes \, M _{n} ( \Psi _{k} \, , ... , \Psi _{k+n-1} ) \otimes 
\Psi _{k+n} \otimes \cdot \! \cdot \! \cdot \otimes \Psi _{m}
\end{align*}
for $n\leq m$, where $(-)^{\sigma _{k} } =(-)^{M_{n} ( \Psi _{1} + \cdots + \Psi _{k-1} )}$ denotes the grading. 
Note that a constant $M_{0}$ yields a map $\bM _{0}$ inserting $M_{0}$ in the tensor products; $(\bM _{0})^{q}$ maps $\cH ^{m}$ to the coefficient of the $q$-degree homogeneous polynomial of $M_{0}$\,. 
By defining $\bM _{n} ( \Psi ^{\otimes m} ) \equiv 0$ for $n > m$, this $\bM _{n}$ gives a coderivation on the tensor algebra, $\bM _{n} : \cT (\cH ) \rightarrow \cT (\cH )$\,. 
The sum $\bM = \sum_{n=0}^{\infty } \bM _{n}$ also gives a coderivation. 
We write $\ld \mathbf{C}_{k} , \mathbf{D}_{l} \rd $ for the graded commutator of two coderivation, 
\begin{align*}
\Ld \, \mathbf{C}_{k} \, , \, \mathbf{D}_{l} \, \Rd \equiv \mathbf{C}_{k} \, \mathbf{D} _{l} - (-)^{\mathbf{C}_{k} \mathbf{D}_{l} } \mathbf{D}_{l} \, \mathbf{C}_{k} \, ,   
\end{align*}
which is a coderivation given by a map from $\cH ^{\otimes m}$ to $\cH ^{\otimes m-k-l+2}$. 
We write $\frac{1}{1-\Psi } \equiv \sum _{n=0}^{\infty } \Psi ^{\otimes n}$ for the group-like element of the tensor algebra $\cT (\cH )$, where $\cH ^{\otimes 0} = \mathbb{C}$ and $\Psi ^{\otimes 0} = 1$. 
A natural projection $\pi _{n} : \cT (\cH ) \rightarrow \cH ^{\otimes n}$ is defined by $\pi _{n} : \sum_{k =0}^{\infty } \Psi _{1} \otimes \cdot \! \cdot \! \cdot \otimes \Psi _{k} \longmapsto \Psi _{1} \otimes \cdot \! \cdot \! \cdot \otimes \Psi _{n}$\,. 
When a coderivation $\bM $ is nilpotent, a pair $(\cT (\cH ) , \bM )$ defines an (weak) $A_{\infty }$ algebra. 
For each $\cH ^{\otimes n}$, the $A_{\infty }$ relation $\bM ^{2} = 0$ of $\bM = \sum_{n=1}^{\infty } \bM _{n}$ can be written as follows 
\begin{align*}
\pi _{n} \, \sum_{k=0}^{n-1} \Ld \, \bM_{k+1} \, , \bM_{n-k} \, \Rd \frac{1}{1-\Psi } 
= \sum_{k=0}^{n-1} M_{k+1} \big{(} \overbrace{\Psi , ... , \Psi }^{k} , M_{n-k} (\overbrace{\Psi , ... , \Psi }^{n-k} ) , \overbrace{\Psi , ... , \Psi }^{n-k-1} \big{)} = 0 \, . 
\end{align*}
Furthermore, with a coderivation $\bLambda _{-g,p}$ inserting $\Lambda _{-g,p}$ into the Fock space $\cT (\cH )$, we get
\begin{align*} 
\pi _{1} \, \Ld \, \bM \, , \, \bLambda _{-g,p} \, \Rd \frac{1}{1-\Psi } = \sum_{n=1}^{\infty } \sum_{k=0}^{n-1} M_{n} \big{(} \overbrace{\Psi , ... , \Psi }^{k} , \Lambda _{-g,p} , \overbrace{\Psi , ... , \Psi }^{n-k-1} \big{)} \, . 
\end{align*}
The suspension of $M_{n}^{\prime }$ is given by $M_{n} ( \Psi _{1} , ... , \Psi _{n} ) \equiv (-)^{\sum_{k=1}^{n} (n-k) ( \textrm{gh} [\Psi _{k} ] + s [\Psi _{k} ] +1 ) } M_{n}^{\prime } ( \Psi _{1} , ... , \Psi _{n} )$\,. 
Then, we find 
$\la \Psi _{0}, \, M_{n}( \Psi _{1} \dots , \Psi _{n}) \ra = - (-)^{\epsilon [ \Psi _{0} ] } \la M_{n}(\Psi _{0} , \Psi _{1} \dots , \Psi _{n-1} ) , \, \Psi _{n} \ra $ if $\bM$ is cyclic. 
For a derivation $\partial _{s}$, we find $\partial _{s} M_{n} ( \Psi _{1} , ... , \Psi _{n} ) = \sum_{i=0}^{n-1} (-)^{\sigma _{i}} M_{n} ( \Psi _{1} , ... , \partial _{s} \Psi _{i+1} , ... , \Psi _{n} )$, where the sign is defined by $\sigma _{i} = \partial _{s} ( \epsilon [\Psi _{1} ] + \cdots + \epsilon [\Psi _{i}] +1 )$.

\end{document}